\documentclass[conference]{IEEEtran}
\IEEEoverridecommandlockouts
\usepackage{amsmath}
\usepackage{amssymb}
\usepackage{algorithm}
\usepackage{subcaption}
\usepackage{algpseudocode}
\usepackage{listings}
\usepackage{amsthm}
\usepackage{booktabs}
\usepackage{tabularx}
\usepackage{multirow}
\usepackage{graphicx}

\usepackage{xifthen}
\usepackage{mathtools}
\usepackage{bm}
\usepackage{xparse}
\usepackage[pagebackref=true,breaklinks=true,letterpaper=true,colorlinks,bookmarks=false]{hyperref}
\usepackage{soul}

\NewDocumentCommand{\vect}{ O{} O{} m }{\bm{#3}\ifthenelse{\isempty{#1}}{}{^{(#1)}}\ifthenelse{\isempty{#2}}{}{_{#2}}}

\NewDocumentCommand{\mat}{ O{} O{} m }{\bm{#3}\ifthenelse{\isempty{#1}}{}{^{(#1)}}\ifthenelse{\isempty{#2}}{}{_{#2}}}

\NewDocumentCommand{\ten}{ O{} O{} m }{\bm{\mathcal{#3}}\ifthenelse{\isempty{#1}}{}{^{(#1)}}\ifthenelse{\isempty{#2}}{}{_{#2}}}

\newcommand{\pr}{p_r}
\newcommand{\pc}{p_c}
\newcommand{\subsubsubsection}[1]{\paragraph{#1}}

\def\BibTeX{{\rm B\kern-.05em{\sc i\kern-.025em b}\kern-.08em
    T\kern-.1667em\lower.7ex\hbox{E}\kern-.125em}}
    
\begin{document}

\title{Distributed Non-Negative Tensor Train Decomposition
}

\author{\IEEEauthorblockN{Manish Bhattarai}
\IEEEauthorblockA{\textit{Theoretical Division}\\
\textit{Los Alamos National Laboratory}\\
Los Alamos, NM, USA\\
ceodspspectrum@lanl.gov}
\and
\IEEEauthorblockN{Gopinath Chennupati}
\IEEEauthorblockA{\textit{Information Sciences}\\
\textit{Los Alamos National Laboratory}\\
Los Alamos, NM, USA\\
gchennupati@lanl.gov}
\and
\IEEEauthorblockN{Erik Skau}
\IEEEauthorblockA{\textit{Information Sciences}\\
\textit{Los Alamos National Laboratory}\\
Los Alamos, NM, USA\\
ewskau@lanl.gov}
\and
\IEEEauthorblockN{Raviteja Vangara}
\IEEEauthorblockA{\textit{Theoretical Division}\\
\textit{Los Alamos National Laboratory}\\
Los Alamos, NM, USA\\
rvangara@lanl.gov}
\and
\IEEEauthorblockN{Hristo Djidjev}
\IEEEauthorblockA{\textit{Information Sciences}\\
\textit{Los Alamos National Laboratory}\\
Los Alamos, NM, USA \\
djidjev@lanl.gov}
\and
\IEEEauthorblockN{Boian S. Alexandrov}
\IEEEauthorblockA{\textit{Theoretical Division}\\
\textit{Los Alamos National Laboratory}\\
Los Alamos, NM, USA \\
boian@lanl.gov}
}

\maketitle

\begin{abstract}
The era of exascale computing opens new venues for innovations and discoveries in many scientific, engineering, and commercial fields. However, with the exaflops also come the extra-large high-dimensional data generated by high-performance computing. High-dimensional data is presented as multidimensional arrays, aka tensors. The presence of latent (not directly observable) structures in the tensor allows a unique representation and compression of the data by classical tensor factorization techniques. However, the classical tensor methods are not always stable or they can be exponential in their memory requirements, which makes them not suitable for high-dimensional tensors. Tensor train (TT) is a state-of-the-art tensor network introduced for factorization of high-dimensional tensors. TT transforms the initial high-dimensional tensor in a network of three-dimensional tensors that requires only a linear storage. Many real-world data, such as, density, temperature, population, probability, etc., are non-negative and for an easy interpretation, the algorithms preserving non-negativity are preferred. Here, we introduce a distributed non-negative tensor-train and demonstrate its scalability and the compression on synthetic and real-world big datasets. 
\end{abstract}


\section{Introduction}
Extra-large volumes of data are constantly being generated nowadays in areas such as personalized medicine, biology, space, nuclear science, climate and in many other fields. A common way to store and use such data is to first reduce its size without losing important information. Unfortunately, the classical compression techniques that are searching for repeated patterns often cannot be used when the data comes from high-performance computing (HPC) simulations that require a relatively high-precision and need a high accuracy of prediction. Such simulation data is usually high-dimensional and  is naturally represented by tensors, i.e., as multi-dimensional arrays. Tensors in such applications typically represent multiple concurrent \textit{latent} (not-directly observable) processes (e.g., explosions, earth-quakes, etc.) of the simulated phenomenon, imprinting their signatures in various simulated variables (such as temperature, pressure, density, etc.) in different dimensions (e.g., space/time). 
Despite the differences, most of the simulated phenomena share the property that the information content of the generated data is quite low, i.e., it can be represented by a low number of parameters, which are however not known a priori and are implicit in the data.
Classical tensor decomposition techniques such as Tucker Decomposition \cite{tucker1966some} and Canonical Polyadic Decompositions (CPD) \cite{hitchcock1927expression}, can extract the latent structures and parameters describing the data, which allows a new type of compression.
For instance, if we consider a $d$-dimensional tensor with $n$ elements in each dimension and a \textit{tensor rank} $r$, CPD allows the representation with the smallest number of parameters, $\mathcal{O}(dnr)$, among all tensor decomposition methods. Unfortunately, finding the canonical rank is an NP-hard problem \cite{haastad1990tensor} and, moreover, the approximation with a fixed rank can be ill-posed \cite{de2008tensor}. For the same tensor, the Tucker decomposition is stable and will have $\mathcal{O}(dnr+r^d)$ number of parameters. Tucker decomposition was previously proposed for compression of large scientific data and it demonstrated an excellent compressibility \cite{austin2016parallel}.  However, for  tensors  of  a  higher  dimension, $d$ (say 10 or more), Tucker decomposition is not feasible since the memory and amount of operations grow exponentially with the tensor dimension $d$. 
\begin{figure}[htp]
    \centering
    \raisebox{0.4cm}{\includegraphics[width=0.22\linewidth]{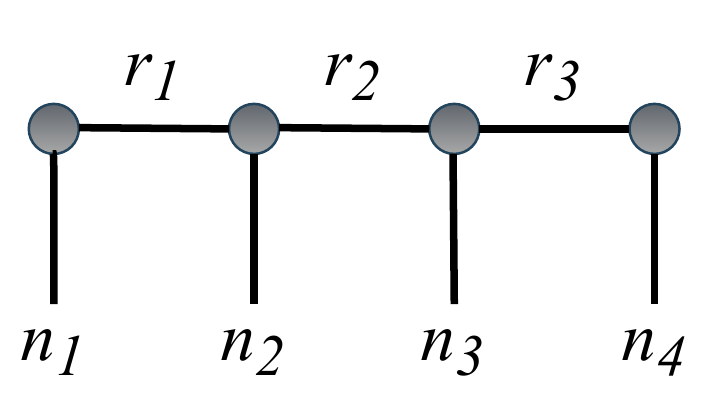}}\hfill
    \includegraphics[width=0.7\linewidth]{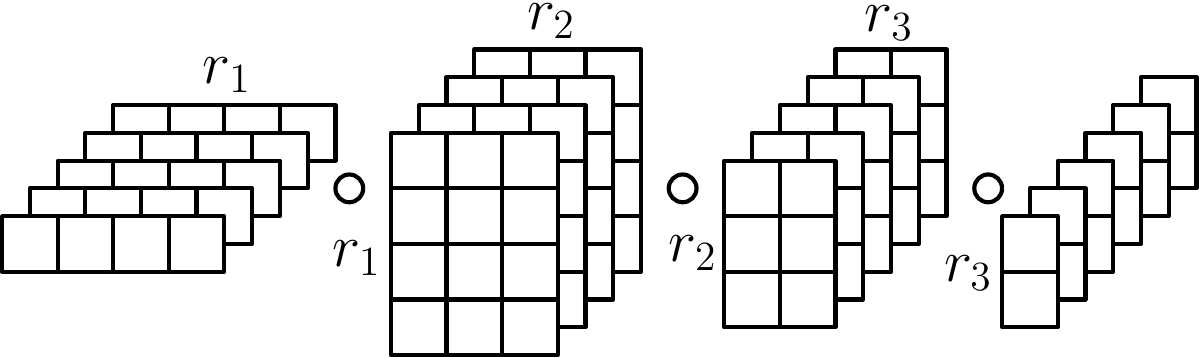}
    \caption{(left) A graphical representation of a TT for a 4-dimensional tensor. A core is represented by a circle, while arms indicate the modes of the tensor and rank indices. The first and last TT cores are matrices, whereas the rest are 3-dimensional tensors. (right) Example of a TT for a $5\times 4\times 5\times 6$ tensor with \textit{ tensor train rank} ($r_1=4, r_2=3, r_3=2$).}
    \label{fig:tensors}
\end{figure}

\textit{Tensor train} (TT) has been introduced in \cite{oseledets1} as a method for decomposition of extra large and high-dimensional tensors targeting application in the solution of partial differential equations. TT is a state-of-the-art tensor network based on Singular Value Decomposition (SVD) process that parameterizes the initial high-dimensional tensor by a network of three-dimensional tensors \cite{oseledets2011tensor}. Figure~\ref{fig:tensors} represents a TT decomposition for a four-dimensional tensor. TT (also known in other areas as a matrix product state \cite{li2018shortcut}) is stable and requires only a linear storage in $d$ with $\mathcal{O}(dnr^2)$ parameters, making it suitable for compression of high-dimensional tensors \cite{wang2019principal}.

Many real-world types of data, such as density, temperature, population, probability, etc., are non-negative and hence algorithms that preserve the non-negativity are preferred in order to retain the interpretability and meaning of the compressed data. Nonnegative factorization is used as a model for recovering latent structures in such data and it has some additional useful features.
For instance, with interpretation of the elements of the data as conditional probabilities, one can make use of the duality between latent graphical models and tensor networks \cite{ishteva2015tensors,robeva2019duality} to obtain a graphical model representation of the data, thereby allowing suitable algorithms for graphical models to be applied for tensor analysis. Graphical models have many applications in finance, machine learning, computer vision, speech recognition, and bioinformatics. Moreover, in contrast with the general case, the nonnegative best rank approximation of a tensor \textit{always} exists \cite{lim2009nonnegative} and it is almost always \textit{unique} \cite{qi2016semialgebraic}, which are useful properties for tensor data analysis.

Due to the large input size and the additional computational efforts needed to ensure the nonnegativity of the resulted low-dimensional factors, nonnegative tensor train (nTT) cannot be applied to data coming from many real-life applications, unless massive parallelism is used. Unfortunately, no distributed algorithm for computing an nTT decomposition has been designed yet.
In this paper, we introduce such distributed  nTT algorithm 
and analyze its performance.
Our algorithm is based on distributed implementation of nonnegative matrix factorization (NMF) and tensor unfolding and reshaping operations. In the next sections, we describe details of the algorithm, its implementation, and analyze its compression ability and scalability on synthetic and real-world datasets.

\section{Background}
\begin{figure}[h]
    \centering
    \includegraphics[width=\linewidth]{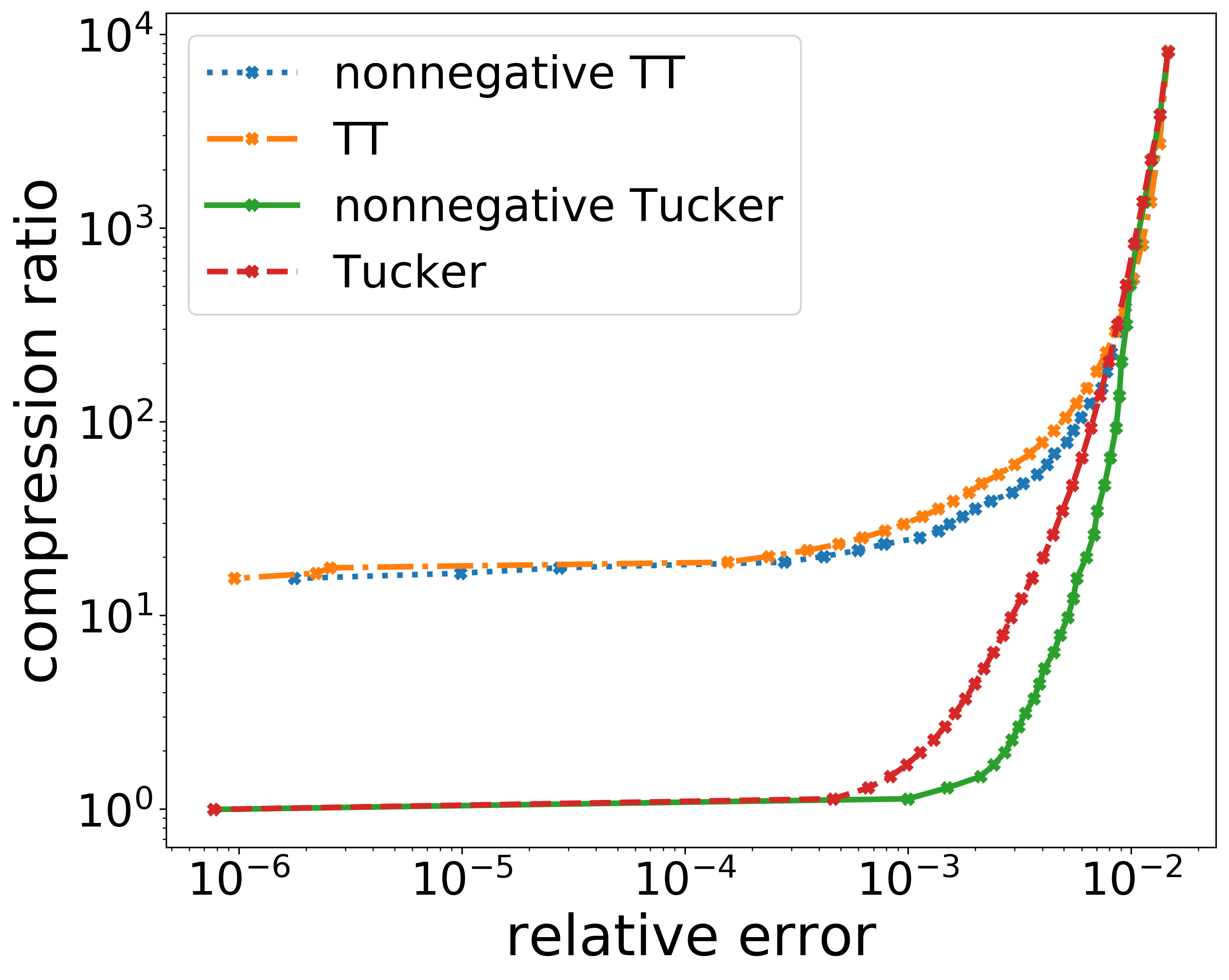}
    \caption{Compression versus relative error for various algorithms on a synthetic data of dimensions 32 x 32 x 32 x 32.}
    \label{fig:Compr-Tuck-TT}
\end{figure}

Tensor train (TT) networks were first introduced by Oseledets in \cite{oseledets1}. 
Unlike other tensor decompositions, TT, being based on a sequence of SVD operations, is easy to compute and does not suffer from the curse of dimensionality. Based on Eckart–Young theorem \cite{eckart1936approximation}, SVD can be used to provide the best low-rank approximation; hence, we expect that TT (which ignores the nonnegativity) will lead to excellent compression results. From the number of parameters that Tucker decomposition and TT require, we can also conclude that the compresibility of TT will be better that the compresibility of Tucker decomposition. In fact, Figure~\ref{fig:Compr-Tuck-TT} shows this phenomenon, where we present the compression versus relative error results for TT, nTT, Tucker and nonnegative Tucker decompositions, performed for the same synthetically generated tensor calculated with~\cite{novikov2020tensor, kossaifi2019tensorly}. Importantly, although with the smallest number of parameters, CP decomposition uses a single rank for all modes, which makes it less robust compared to TT and Tucker decompositions. 

The main factor to be taken into consideration when computing tensor train decompositions is the tradeoff between relative error and the tensor train rank \cite{ding2020tensor}. 
Decompositions of specified ranks could have either a large computational complexity, or may result in a large error. A number of algorithms to compute tensor trains exists  (each addressing this issue in a different way), such as, the sequential approach known as TT-SVD \cite{oseledets2011tensor}, Alternating Least Squares (ALS) algorithms \cite{holtz2012alternating}, and Density Matrix Renormalization Group (DMRG) algorithms from quantum physics \cite{huckle2013computations,marti2010complete}.
 
 A number of distributed TT algorithms \cite{fonal2019distributed}, \cite{wang2019distributed}, \cite{wang2020adtt} exist in the literature. An application of a new out-of-memory algorithm for truncated SVD computation for use in the TT algorithms, targeting tensors so large that they cannot fit in the memory, was also proposed in \cite{9110583}. Recently, a MapReduce-based distributed TT algorithm for large-scale dimension reduction and classification  was introduced in \cite{fonal2019distributed}. However, the proposed framework is unable to achieve significant scalability compared to non-distributed implementation. Next,~\cite{wang2019distributed} used a blocking strategy for distributed TT, where the reshape operation of the tensor block is done locally before decomposition.  They compute the relative error between the TT cores calculated in a distributed manner and the non-distributed approach. In such cases, the paper fails to evaluate the correctness of the framework for large-sized tensors where the non-distributed approach cannot be applied. Recently, \cite{wang2020adtt} proposed a distributed TT for processing Internet of things (IoT) data. Similar to \cite{wang2019distributed}, this framework has the limitations on the ability to compute the final reconstruction error from the factors of a large scale tensor.  

Nonnegative tensor train is a much less explored topic, 
with only a handful of published works \cite{lee2016nonnegative, shcherbakova2019nonnegative, shcherbakova2019nonnegative2}.

In \cite{lee2016nonnegative}, Lee \textit{et. al.} proposed a nonnegative tensor train and provide an NTT-HALS algorithm to compute it. NTT-HALS aims to minimize the Frobenius norm reconstruction error for a given tensor train rank by using a Hierarchical Alternating Least Squares (HALS) procedure. The authors demonstrated that the storage cost of their  NTT-HALS decomposition is significantly lower than a corresponding nonnegative Tucker decomposition. Due to its computational complexity, the NTT-HALS algorithm is ill suited to large high-dimensional tensors. In \cite{shcherbakova2019nonnegative2} the authors propose an algorithm called NTTF, which relies on successive unfoldings and NMF approximations. That algorithm can accommodate a desired relative error by ensuring that each NMF approximation achieves a certain accuracy. This is computationally efficient approach, but it can result in unbalanced and large tensor train ranks. Finally, in \cite{shcherbakova2019nonnegative}, Shcherbakova proposes a NTT-MU algorithm similar in goal to \cite{lee2016nonnegative}, but based on a DMRG algorithm to additionally minimize the tensor train ranks. DMRG-based algorithms simultaneously optimize over consecutive pairs of tensors in the train, this allows the tensor train rank between the pair to be adjusted. However, the proposed algorithm utilizes multiplicative update to ensure nonnegativity, which suffers from the inability to converge to a high accuracy.

\begin{table}[htp]\footnotesize
	\centering
	\caption{Notation} \label{tab:notations}
	\begin{tabular}{|c|c|l|}
		\toprule
		\textbf{Notation} & \textbf{Dimensions} & \textbf{Description} \\
		\midrule
		$\ten{A}$ & $n_1 \times \hdots \times n_{d}$ & Tensor of interest\\
		$\ten[i][]{G}$ & $r_{i-1} \times n_i \times r_i$ & Tensor train core\\
		$p_i$ & scalar & Processor count along each dimension \\
		$\ten[i_1,\dots,i_d]{A}$ & $\frac{n_1}{p_1} \times \hdots \times \frac{n_{d}}{p_{d}}$ & Distributed block of the tensor  \\
		$m$   & scalar & Number of rows of a matrix \\
		$n$   & scalar & Number of columns of a matrix \\
		$\mat{X}$ & $m \times n$ & Input matrix to NMF \\
		$\pr$ & scalar & Proc. count along rows of matrix\\
		$\pc$ & scalar & Proc. count along columns of matrix\\
		$\mat[i,j]{X}$ & $\frac{m}{p_r} \times \frac{n}{p_c}$ & Distributed block of matrix \\
		$r$   & scalar & Low rank \\
		$\mat{W}$ & $m \times r$ & Left low rank factor \\
		$\mat{H}$ & $r \times n$ & Right low rank factor \\
		$p$ & scalar & Total processor count, $p = \prod_{i=1}^{d} p_i $ \\
		$\mat{(W^i)^{j}}$ & $\frac{m}{p} \times r$ & Left low rank factor on $(i,j)^{th}$ \\
		&&processor\\
		$\mat{(H^j)^{i}}$ & $r \times \frac{n}{p}$ & Right low rank factor on $(i,j)^{th}$  \\
		&&processor\\		
		
		$\mat[i]{W}$ & $\frac{m}{p_r} \times r$ & Block of factor W  \\
		&& corresponding  to block $\mat[i,j][]{X}$ \\
		$\mat[j]{H}$ & $r \times \frac{n}{p_c}$ & Block of factor H  \\
		&& corresponding to block $\mat[i,j][]{X}$ \\ 
		\bottomrule
	\end{tabular}
\end{table}

\section{Algorithmic details}
The tensor-train format is an efficient representation of a higher dimensional tensor in  terms of storage requirements and computational robustness \cite{oseledets1},\cite{oseledets2}. Tensor train representation achieves such performance metrics by combining the major advantages of the  Canonical format  and the Tucker format \cite{kolda2009tensor}. Tensor train decomposes a $d$-dimensional tensor $\ten{A}\in \mathbb{R}^{n_1 \times \hdots \times n_d}$ into \textit{d} 3-dimensional tensors $\ten[i]{G} \in \mathbb{R}^{r_{i-1}\times n_{i}\times r_{i}}$, where $r_{0} = r_{d} = 1$ (so $\ten[1]{G}$ and $\ten[d]{G}$ are actually matrices) and $r_{k} \geq 1$ for $k = 1,...,d-1$,  such that


\begin{equation}
\begin{aligned}
    \ten{A} = \ten[1]{G} \circ \ten[2]{G} \circ \hdots \circ \ten[d]{G},
    \end{aligned}
\end{equation}
where $\ten{A} \circ \ten{B}$ is used to define tensor-tensor multiplication with contraction along the last axis of $\ten{A}$ and the first axis of $\ten{B}$, 
\linebreak
$(\ten{A} \circ \ten{B})_{i_1,i_2,\hdots,i_{d-1},j_2,j_3,\hdots,j_d} = \sum_{k} \ten[][i_1,\hdots,i_{d-1},k]{A} \ten[][k,j_1,j_2,\hdots,j_d]{B}$. Here, the 3-dimensional tensors $\ten[i]{G}$ are called  \textit{TT cores} and the numbers $r_{1},\dots,r_{d-1}$ are called  \textit{TT ranks}. 
Based on the above, any element of the tensor $\ten{A}$ can be further be represented as

\begin{equation}
\begin{aligned}
\ten[][i_{1},\dots,i_{d}]{A} = \sum_{k_1, \hdots, k_{d-1}}^{r_1, \hdots, r_{d-1}} \ten[1][i_1,k_1]{G} \ten[2][k_1,i_2,k_2]{G} \hdots \ten[d][k_{d-1},i_d]{G}.
\end{aligned}
\end{equation}

Table~\ref{tab:notations} shows the notations used in the paper.

\begin{figure}[htp]
    \centering
    \includegraphics[scale=0.6]{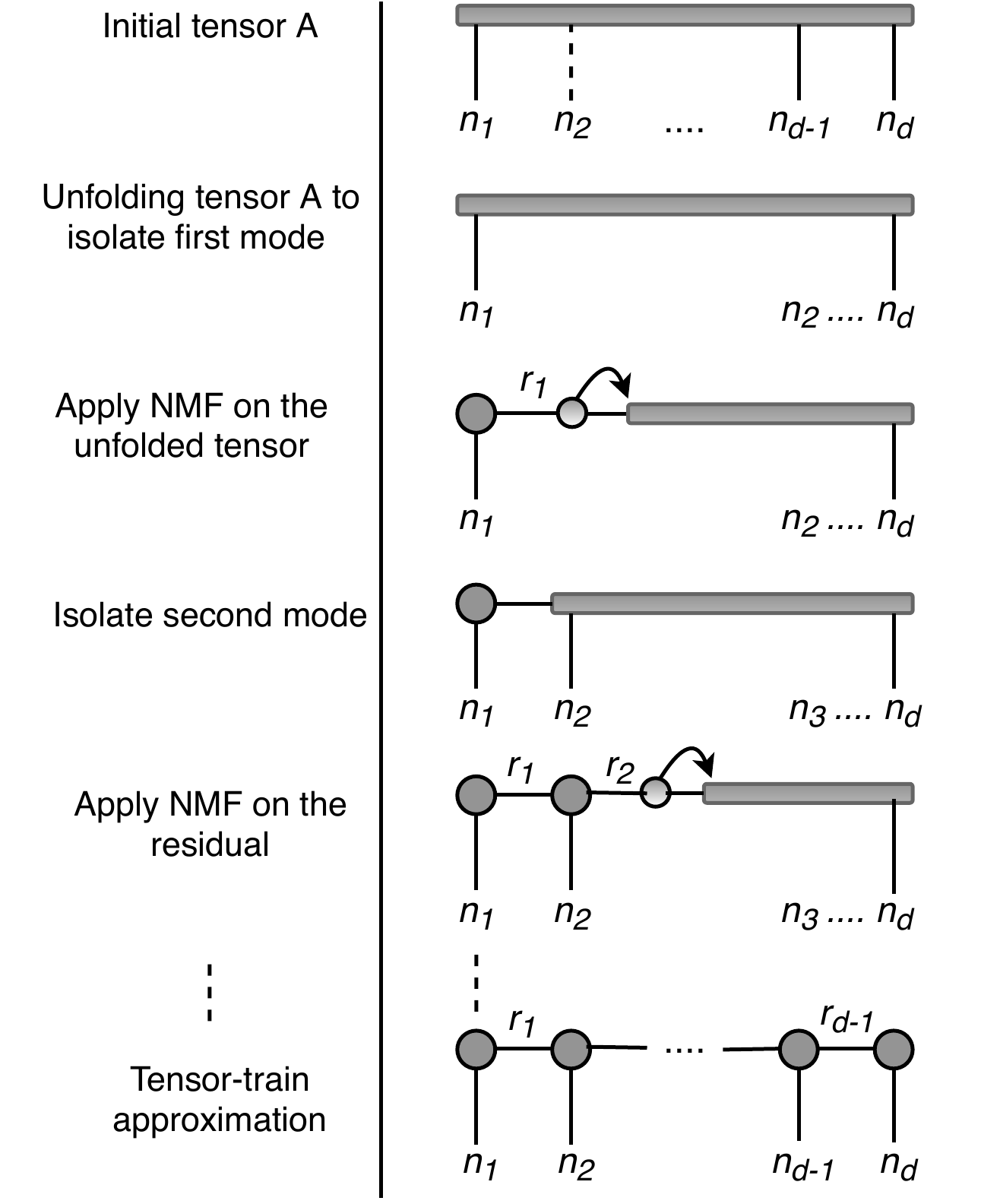}
    \caption{The decomposition procedure of a tensor into a TT format iteratively isolates the next mode and applies NMF to get the next factor. The TT cores are represented by circles, while the remaining unfactored matrix is represented by a bar.}
    \label{fig:illustration}
\end{figure}

\subsection{Non-Negative Tensor Train}

\begin{algorithm}[htp]
    \caption{$\mat[i,j]{X}$ = $\operatorname{distReshape} (\ten[i_i,\hdots,i_d]{A},[m,n], [p_{i},p_{j}])$ -- Distributed reshaping of tensor $\ten{A}$ into $\mat{X}$}. \label{alg:distreshape}
    \begin{algorithmic}[1] 
    \Require: Tensor $\ten[i_1,i_2,\hdots,i_d]{A}\in \mathbb{R}^{\frac{n_{1}}{p_{1}} \times \hdots \times \frac{n_{d}}{p_d}}$,target matrix shape $m \times n$ and target processor grid size $p_{r} \times p_{c}$.
    \State  Each MPI rank writes a block of $\ten[i_1,i_2,\hdots,i_d]{A}$.
    \State Perform global reshaping of the tensor from tensor dimensions $n_{1} \times n_{2} \times \hdots \times n_{d}$ into $m \times n$  i.e $\ten{X}=reshape(\ten{A}, [m_{1},m_{2},...m_{k}]$). 
    \State Compute in-memory data $\mat[i,j]{X}$ for each MPI rank from the reshaped tensor $\mat{X}$ where size of $\mat[i,j]{X} = \frac{m}{p_r} \times \frac{n}{p_c}$ 
    \Ensure  $\mat{X} \in \mathbb{R}^{ m \times n }$ and $\mat[i,j]{X} \in \mathbb{R}^{\frac{m}{p_r} \times \frac{n}{p_c}}$
     \end{algorithmic}   
\end{algorithm}

\begin{algorithm}[htp]
    \caption{$\ten[][1]{G}, \ten[][2]{G}, \hdots, \ten[][d]{G}$ = $\operatorname{distnTT}(\ten[i_1,i_2,\hdots,i_d][]{A},\epsilon)$ -- Distributed non negative tensor train algorithm}\label{alg:distTT}
	\begin{algorithmic}[1]
    \Require  Tensor $\ten[i_1,i_2,\hdots,i_d]{A}\in \mathbb{R}^{\frac{n_{1}}{p_{1}} \times \hdots \times \frac{n_{d}}{p_d}}$ and threshold $\epsilon$.
    \State Set $r_{0}$ = $r_{d}$ = 1
    \State Set $\mat[i,j]{A} = \operatorname{distReshape}$ $(\ten[i_1,i_2,\hdots,i_d]{A},[n_1,\frac{S }{n_1}],[p_1,p/p_1])$  \Comment{$S=n_1 \times n_2 \hdots \times n_d$ and $p=p_1 \times p_2 \hdots \times p_d$}
    \For{$l$ in 1 to $d-1$}
    \State $\mat[i,j]{X} = \operatorname{distReshape}$ $(\mat[i,j]{A},[r_{l-1} n_{l},\frac{S}{r_{l-1} n_{l}}],[p_1,p/p_1])$ \Comment{$S=n_l \times n_{l+1} \hdots \times n_d$}
    \State $\mat[i,j]{U} \mat{\Sigma} \mat{V^{(i,j)T}}$ =  $\operatorname{distSVD}( \mat[i,j]{X})$  \Comment{$\Sigma \in \mathbb{R}^{N \times N}$}
    \State Choose $r_l$ smallest $k$ such that $\frac{\sqrt{\sigma^{2}_{k+1}+ ... + \sigma^{2}_{N}}}{\sqrt{\sigma^{2}_{1}+ ... + \sigma^{2}_{N}}} \leq \epsilon$ 
    \State $\mat{(W^i)^{j}}$, $\mat{(H^j)^{i}} =\operatorname{distBCDnmf}(\mat[i,j]{X}, r_{l}) $
    \State $\ten{G}^{(l)}=\operatorname{reshape}(\operatorname{all\_gather}(\mat{(W^i)^{j}}), [r_{l-1},n_{l},r_{l}])$ \Comment{Perform global all\_gather on $\mat{(W^i)^{j}}$ along row-wise distribution. $\mathbb{R}^{r_{l-1}\times n_{l}\times r_{l}}$}
    \State $\mat[i,j]{A}=\mat{(H^j)^{i}}$ \Comment{$\in \mathbb{R}^{r_{l} \times n_{l+1} ... n_{d}}$  is 1D-distributed} 
    \EndFor
    \State Set d-th core to $\ten[d]{G}_{:,:,1} = \operatorname{all\_gather}(\mat[i,j]{A})$ \Comment{Perform global all\_gather on $\mat[i,j]{A}$ along column-wise distribution}
    \Ensure Approximation $\ten[1]{G} \circ \ten[2]{G} \circ \hdots \circ \ten[d]{G}$ of $\ten{A}$ in TT format with cores $\ten[1]{G}, \ten[2]{G}, \hdots, \ten[d]{G}$ and TT ranks $r_{0},r_{1},....,r_{d}$
    \end{algorithmic} 
    
\end{algorithm}
Figure~\ref{fig:illustration} illustrates the construction of a $d$-dimensional TT. Given a $d$-dimensional tensor $\ten{A} \in \mathbb{R}^{n_1 \times n_2 \times ... \times n_d}$, the first step of TT is to isolate the first mode of $\ten{A}$ with left unfolding to produce a matrix $\mat{X} \in \mathbb{R}^{n_1 \times  n_2 n_3  \dots n_d}$. NMF is used to reduce $\mat[][]{X}$ to its corresponding factors $\mat{W} \in \mathbb{R}^{n_1 \times r_1}$ and $\mat{H} \in \mathbb{R}^{r_1 \times n_2 n_3  ... n_d}$. The rank of this NMF decomposition, $r_1$, is selected by an SVD-based heuristic, namely, $r_1$ is selected by choosing it to be equal to the smallest $k$ such that $\sqrt{\sigma^{2}_{k+1}+ ... + \sigma^{2}_{N}}/\sqrt{\sigma^{2}_{1}+ ... + \sigma^{2}_{N}} \leq \epsilon$, where $\sigma_{i}$ is the $i^{th}$ singular value of $\mat{X}$ and $N = \min(n_1,n_2 n_3  \dots n_d )$. The singular values decrease and reach a plateau, separating a cluster of non-latent dimensions ($r \geq r_{s+1}$) to that of latent ones ($r \leq r_{s}$). The number of these larger singular values is used to calculate the effective rank of a  matrix \cite{roy2007effective} as well in Bayesian PCA \cite{bishop1999bayesian}. There are other sophisticated approaches to identify latent dimensionality of NMF \cite{distnmfk}, which could be included in our method. The left NMF factor ${W}$ is the first core of the non-negative tensor train decomposition of \ten[][]{X}. The right NMF factor $\mat{H}$ is further reshaped to $\mat{X} \in \mathbb{R}^{r_1n_2 \times n_3 \times n_4 \times... \times n_d}$. The procedure is repeated until the final NMF factor $\mat{H} \in \mathbb{R}^{r_{d-1} \times n_d}$, i.e., core $\ten[d][]{G}$ of nTT is obtained.

\begin{algorithm}[htp]\footnotesize
    \caption{$\mat{(W^i)^{j}}$, $\mat{(H^j)^{i}}$ = $\operatorname{distBCDnmf}(\mat[i,j]{X}, r)$ -- Distributed BCD NMF algorithm}\label{alg:distbcd}
	\begin{algorithmic}[1]
    \Require  $\mat[i,j]{X} \in \mathbb{R}_{+}^{\frac{m}{p_r} \times \frac{n}{p_c}}$ and desired rank $r$.
    \State Initialize $\mat{(W^i)^{j}}$, $\mat{(H^j)^{i}}$ = $\operatorname{rand}(\frac{m}{p}, r)$, $\operatorname{rand}(r, \frac{n}{p})$ 
    \State $\mat{(W^i)^{j}}_m$, $\mat{(H^j)^{i}}_m$ = $\frac{\mat{(W^i)^{j}}}{||\mat[]{W}||}\sqrt{||\mat[]{X}||}$, $\frac{\mat{(H^j)^{i}}}{||\mat[]{H}||}\sqrt{||\mat[]{X}||}$ \Comment{Normalize}
    \State $\mat{{HH^T}}$, $((\mat{{XH^T}})^i)^j$ =$\operatorname{distMM^T}(\mat{(H^j)^{i}_m)}, \operatorname{distXH^T}(\mat[i,j]{X},\mat{(H^j)^{i}_m)}$
    \State $t$, $obj$ = $1.0$, $\frac{1}{2}||X||$ \Comment{Correction of $\mat{W}$ and $\mat{H}$}
    \For{$l$ in $max\_iters$}
        \Statex \textbf{/* Update W given H */}
        \State $\mat{{(W_{m}{HH^T}^i)^j}}$ = $\mat{(W^i)^{j}}_m \mat[]{{HH^T}}$ \Comment{$i^{th}$ proc $\mathbb{R}_{+}^{\frac{m}{p} \times r}$} 
        \State $\mat{(G_{W}^i)^j}$  = $\mat{{(W_{m}{HH^T}^i)^j}}$ -- $((\mat{{XH^T}})^i)^j$ \Comment{$\mat{(G_{W}^i)^j} \in \mathbb{R}_{+}^{\frac{m}{p} \times r}$}
        \State $\mat{(W^i)^{j}}$ = $\operatorname{max}(0, \frac{\mat{(W^i)^{j}_m}}{||\mat{{HH^T}}||})$
        \State $\mat{(W^i)^{j}} /= ||\mat{W}||_1$ \Comment{Normalize with $L_1$ norm }
        \State $\mat{W^TW}$ = $\operatorname{distMM^T}(\mat{((W^i)^{j}})^T)$
        \Statex \textbf{/* Update H given W*/}
        \State $\mat{(W^TWH_m^j)^i}$ = $\mat{W^TW} \mat{(H^j)^{i}}_m$ \Comment{$\mat{(W^TWH_m^j)^i} \in \mathbb{R}_{+}^{r \times \frac{n}{p} }$}
        \State $((\mat{{W^TX}})^j)^i$ = $\operatorname{distWTX}(\mat[i,j]{X}, \mat{(W^i)^{j}})$
         \State $\mat{(G_H^j)^i}$  = $\mat{(W^TWH_m^j)^i}$ -- $((\mat{{W^TX}})^j)^i$ \Comment{$\mat{(G_H^j)^i} \in \mathbb{R}_{+}^{r \times \frac{n}{p}}$}
        \State $\mat{(H^j)^{i}}$ = $\operatorname{max}(0,\mat{(H^j)^{i}_m} - \frac{\mat{(G_H^j)^i}}{ \sqrt{WTW}})$
        \State $\mat{{HH^T}}$ = $\operatorname{distMM^T}(\mat{(H^j)^{i}})$
        \State $((\mat{{XH^T}})^i)^j$ = $\operatorname{distXH^T}(\mat[i,j]{X},\mat{(H^j)^{i}})$
        \If{$\frac{1}{2}||\mat{X}-\mat{WH}||^2 >= obj$} \Comment{\textbf{/* Correction */}}\label{bcd:start_correct}
            \State $Initialize \; \mat{(W^i)^{j}}$ and $\mat{(H^j)^{i}}$
            \State $\mat{{HH^T}}$ = $\operatorname{distMM^T}(\mat{(H^j)^{i}})$
            \State $((\mat{{XH^T}})^i)^j$ = $\operatorname{distXH^T}(\mat[i,j]{X},\mat{(H^j)^{i}})$ \label{bcd:end_correct}
        \Else{} \Comment{\textbf{/* Extrapolation */}}\label{bcd:start_extrapolate}
            \State $w$ = $(t-1)/\frac{1+\sqrt{1+4t^2}}{2}$
            \State $w_W$ = $\operatorname{min}(w, \delta \sqrt{||\mat{HH^T_{(l-1)})}||/||\mat{{HH^T}}||})$
            \State $w_H$ = $\operatorname{min}{w, \delta \sqrt{||\mat{W^TW_{(l-1)}}||/||\mat{{W^TW}}||}}$
            \State $\mat{(W^i)^{j}}_m$ = $\mat{(W^i)^{j}}$ + $w_W$*($\mat{(W^i)^{j}}$--$\mat{(W^i)^{j}_{l-1}}$)
            \State $\mat{(H^j)^{i}}_m$ = $\mat{(H^j)^{i}}$ + $w_H$*($\mat{(H^j)^{i}}$--$\mat{(H^j)^{i}_{l-1}}$) 
            \State  $t,\; obj$ = $\frac{1+\sqrt{1+4t^2}}{2},\; \frac{1}{2}||\mat{X}-\mat{WH}||^2$ \label{bcd:end_extrapolate}
        \EndIf
    \EndFor
    \State \Return $\mat{(W^i)^{j}}$, $\mat{(H^j)^{i}}$
    \Ensure $\mat{(W^i)^{j}} \in \mathbb{R}_{+}^{\frac{m}{p} \times r}$ and $\mat{(H^j)^{i}} \in \mathbb{R}_{+}^{r \times \frac{n}{p}}$ and $\mat{W},\mat{H} \approx \operatorname*{argmin}_{\widetilde{\mat{W}} \geq 0,\widetilde{\mat{H}} \geq 0} ||\mat{X}-\widetilde{\mat{W}} \widetilde{\mat{H}}||_{F}^{2}$
    \end{algorithmic} 
\end{algorithm}

\begin{figure*}
    \centering
    \includegraphics[width=\textwidth]{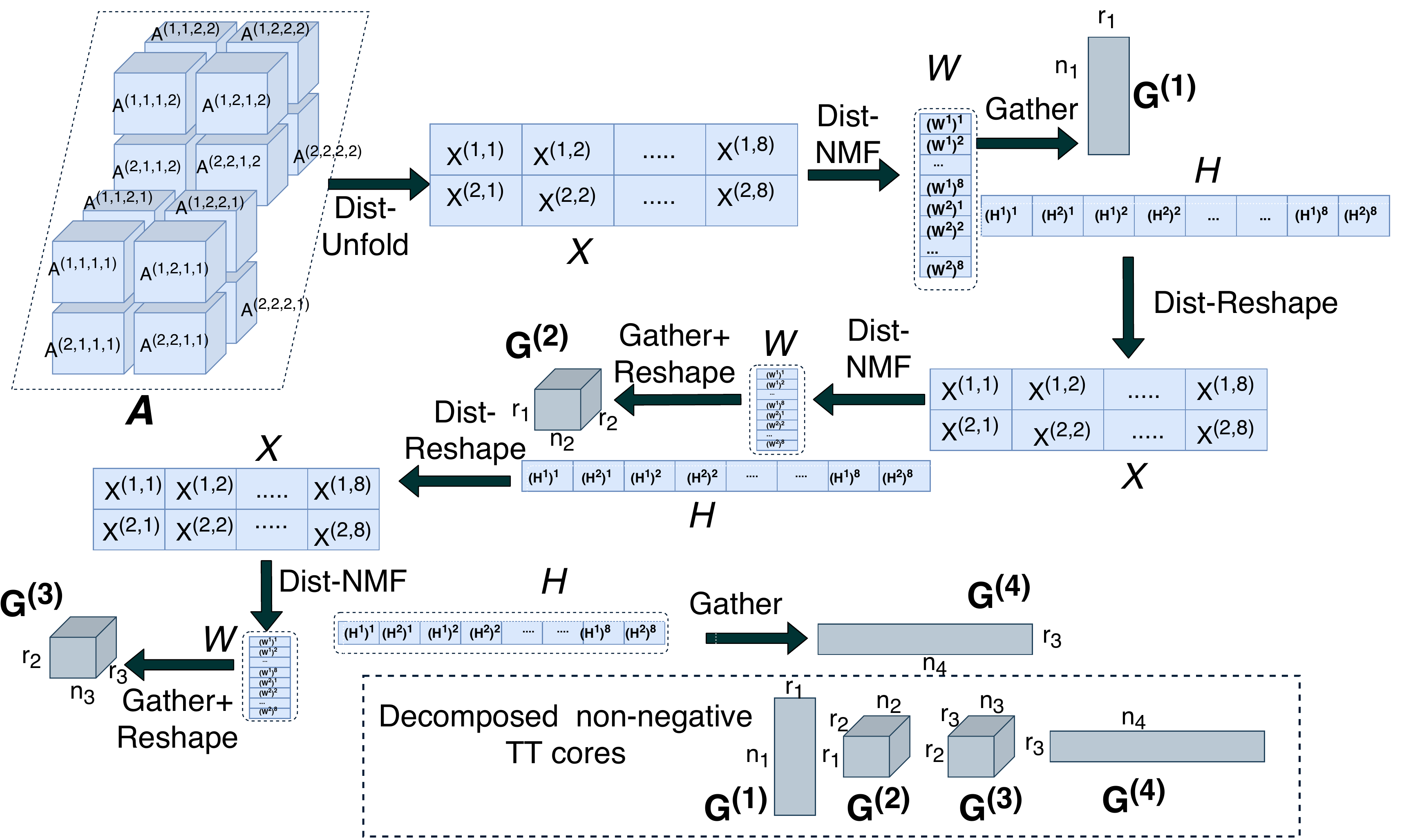}
    \caption{Overview of distributed tensor-train decomposition of a 4D tensor with a processor grid of size 2x2x2x2.}
    \label{fig:overviewl}
\end{figure*}

\subsection{Distribution Strategy}
Figure~\ref{fig:overviewl} shows the distributed TT for a 4-dimensional tensor $\ten{A}$ into four cores $\ten[1][]{G}$, $\ten[2][]{G}$ , $\ten[3][]{G}$ and  $\ten[4][]{G}$. We choose a 4-dimensional processor grid of size $2\times 2\times 2\times 2$ that divides each mode of the tensor. Considering the size of tensor $\ten{A}$ is $n_{1}\times n_{2}\times n_{3}\times n_{4}$, each distributed block of tensor $\ten[i_1,i_2,i_3,i_4][]{A}$ will have a size of  $\frac{n_{1}}{2} \times \frac{n_{2}}{2} \times \frac{n_{3}}{2} \times \frac{n_{4}}{2}$. We first  perform a distributed reshaping/unfolding of the tensor $\ten{A}$ into a matrix $\mat{X}$ of size $n_{1} \times n_{2}n_{3}n_{4}$ using Zarr and Dask packages. We use a Zarr shared file system to store the tensor and the intermediate factors. Dask operates on the Zarr file object for reshaping. Dask first performs global tensor reshape operation via a lazy evaluation/call-by-need approach and then each MPI rank computes the in-memory chunk of data afterward.  The algorithm for distributed reshape is presented in Algorithm~\ref{alg:distreshape}. Each block of reshaped  matrix $\mat{X}$ is of size $\frac{n_{1}}{2} \times \frac{n_{2}}{2} \frac{n_{3}}{2} \frac{n_{4}}{2}$. Next, we carry out the distributed SVD to find the TT rank. Then, we perform the distributed NMF algorithm with block co-ordinate descent (BCD) optimization \cite{xu2013block} presented in Algorithm \ref{alg:distbcd} on the distributed matrix $\mat{X}$. The distribution strategy of the matrix and factors for the distributed NMF is following  \cite{distnmfk}. The dist-NMF produces distributed factors matrices $\mat{W}$ and $\mat{H}$. The matrix $\mat{W}$ is gathered across the processor cores to obtain the first core tensor $\ten[1][]{G}$. Next, the \mbox{1-d} distributed factor matrix $\mat{H}$ undergoes a distributed reshape as per Algorithm~\ref{alg:distreshape} to obtain a 2D-distributed matrix $\mat{X}$. The decomposition is carried out as described above, followed by obtaining factors $\mat{W}$ and $\mat{H}$, which are reshaped accordingly, to obtain cores $\ten[i][]{G}$ and matrices $\mat{X}$, respectively, until the final decomposition is constructed.

Algorithm~\ref{alg:distTT} describes the distributed nonnegative tensor train. Along with the distributed reshape and the SVD, the next key component is the distributed BCD algorithm (see Algorithm~\ref{alg:distbcd}). The inputs to $\operatorname{distBCDnmf}$ algorithm are a 2D distributed input matrix ($\mat[i,j]{X})$ and the rank value ($r$). The algorithm returns intermediate factors, $\mat{(W^i)^j}$ and $\mat{(H^j)^i}$, where the second matrix is used in the reshape operation of the TT algorithm for further decomposition until the last mode of the tensor is processed. In order to get optimal intermediate factors, we employed alternating nonnegative least squares (ANLS) strategy in BCD. This alternates updating each factor while fixing the other factor to be constant. 

\begin{algorithm}[htp]
    \caption{$\mat{{MM^T}}$ = $\operatorname{distMM^T}(\mat{(M^i)^{j}})$ -- Distributed Gram calculation of $\mat{(M^i)^{j}}$}\label{alg:distMMt}
	\begin{algorithmic}[1]
    \Require  $\mat{(M^i)^{j}}$ or $\mat{((M^i)^{j}})^T$ 
    \Statex \textbf{/* To calculate $\mat{{HH^T}}$ or $\mat{{W^TW}}$ */}
    \State $\mat[i,j]{{U}}$ = $\mat{(M^i)^{j}}\mat{((M^i)^{j}})^T$  \Comment{$\mat[i,j]{{U}} \in \mathbb{R}_{+}^{r \times r}$}
    \State $\mat{{MM^T}}$ = $\sum \mat[i,j]{U}$ \Comment{$all\_reduce$ across all proc}
    \Ensure $\mat{{MM^T}} \in \mathbb{R}_{+}^{r \times r}$
    \end{algorithmic} 
\end{algorithm}

\begin{algorithm}[htp]
    \caption{ $((\mat{{XH^T}})^i)^j$ = $\operatorname{distXH^T}(\mat[i,j]{X},\mat{(H^j)^{i}})$ -- Distributed matrix multiplication of $\mat[i,j]{X}$ and $\mat{(H^j)^{i}}$} \label{alg:distXHt}
	\begin{algorithmic}[1]
    \Require  $\mat[i,j]{X} \in \mathbb{R}_{+}^{\frac{m}{p_r} \times \frac{n}{p_c}}$ and  $\mat{(H^j)^{i}} \in \mathbb{R}_{+}^{r \times \frac{n}{p}}$
    \State $\mat[j]{H}$ = $\operatorname{all\_gather}(\mat{(H^j)^{i}})$ \Comment{across processor columns, $\mat[j]{H} \in \mathbb{R}_{+}^{r \times \frac{n}{p_c}}$}
    \State $\mat[i, j]{{V}}$ = $\mat[i, j]{X}\mat{H^{(j)T}}$  \Comment{$\mat[i, j]{{V}} \in \mathbb{R}_{+}^{\frac{m}{p_r} \times r}$}
    \State compute $(\mat{{XH^T}})^i$ = $\sum \mat[i, j]{V}$ \Comment{\textit{reduce-scatter} on processor rows for row-wise distribution}
     \State $(i,j)^{th}$ processor holds $((\mat{{XH^T}})^i)^j$ after \textit{reduce-scatter}
    \Ensure $((\mat{{XH^T}})^i)^j \in \mathbb{R}_{+}^{\frac{m}{p} \times r}$
    \end{algorithmic} 
\end{algorithm}

\begin{algorithm}[htp]
    \caption{$((\mat{{W^TX}})^j)^i$ = $\operatorname{distWTX}(\mat[i,j]{X}, \mat{(W^i)^{j}})$ -- Distributed matrix multiplication of $\mat[i,j]{X}$ and $\mat{(W^i)^{j}}$} \label{alg:distWtX}
    \begin{algorithmic}[1]
        \Require $\mat[i,j]{X} \in \mathbb{R}_{+}^{\frac{m}{p_r} \times \frac{n}{p_c}}$ and $\mat{(W^i)^{j}} \in \mathbb{R}_{+}^{\frac{n}{p} \times r}$
    \State $\mat[i]{W}$ = $\operatorname{all\_gather}(\mat{(W^i)^{j}})$ \Comment{across processor rows, $\mat[i]{W} \in \mathbb{R}_{+}^{ \frac{n}{p_r}} \times r$}
    
        \State compute $\mat[i,j]{Y}$ = $\mat{W^{(i)T}} \mat[i,j]{X}$ \Comment{$\mat[i,j]{Y} \in \mathbb{R}_{+}^{r \times \frac{n}{p_c}}$}
        \State compute  $(\mat{{W^TX}})^j$ = $\sum \mat[i, j]{Y}$ \Comment{\textit{reduce-scatter} on processor columns for column-wise distribution}
          \State $(i,j)^{th}$ processor holds $((\mat{{W^TX}})^j)^i$ after \textit{reduce-scatter}
        \Ensure $ (\mat{{W^TX}})^j)^i \in \mathbb{R}_{+}^{r \times \frac{n}{p}}$
    \end{algorithmic} 
\end{algorithm}

The main computational steps in the distributed BCD are the calculation of matrices $\mat{W^{T}}\mat{W}$ or $\mat{H}\mat{H^{T}}$ (Gram matrices), $\mat{X}\mat{H^{T}}$, and $\mat{W^{T}}\mat{X}$. Each of these computations are being performed multiple times in the BCD algorithm. Therefore, we present the distributed Gram calculation in Algorithm~\ref{alg:distMMt}, Algorithm~\ref{alg:distXHt} computes distributed $\mat{X}\mat{H^{T}}$, and Algorithm~\ref{alg:distWtX} computes distributed $\mat{W^{T}}\mat{X}$. We note the final conditions of BCD try to guarantee convergence to an optimal solution. In this step, the  $\mat{(W^i)^j}$ and $\mat{(H^j)^i}$ matrices are initialized to the initial values (lines~\ref{bcd:start_correct}--\ref{bcd:end_correct} in Algorithm~\ref{alg:distbcd}) when the objective value of optimization is worse than the previous iteration. Otherwise, the intermediate factors are updated (lines~\ref{bcd:start_extrapolate}--\ref{bcd:end_extrapolate} in Algorithm~\ref{alg:distbcd}) accordingly with the use of a user defined hyper parameter ($\delta$). We now evaluate our approach.


\section{Experiments and Results}
\label{sec:experiments}
We run the experiments on the HPC cluster {\em Grizzly}, located at Los Alamos National Laboratory (LANL). {\em Grizzly} has Intel Xeon Broadwell (E5-2695v4) processors with a total of $1490$ compute nodes, where each node has $18$-core dual socket Ivy Bridge processor. Each of the $36$ processors has a clock speed of $2.1$ GHz with a private $L_1$ and $L_2$ caches of sizes $64$KB and $256$KB. Both the sockets share an $L_3$ cache of size $45$MB, where each node contains $128$GB of memory. Grizzly uses Tri-Lab Operating System Stack (TOSS) version 3, while the interconnect is Intel {\em OmniPath} that uses a fat-tree topology. 

Our source code is in Python, where the dependencies include Dask~\footnote{\href{https://docs.dask.org/en/latest/}{https://docs.dask.org/en/latest/}}, Numpy~\footnote{\href{https://numpy.org/doc/1.18/reference/index.html}{https://numpy.org/doc/1.18/reference/index.html}}, MPI4PY~\footnote{\href{https://mpi4py.readthedocs.io/en/stable/}{https://mpi4py.readthedocs.io/en/stable/}}, and Zarr~\footnote{\href{https://zarr.readthedocs.io/en/stable/}{https://zarr.readthedocs.io/en/stable/}}. Our framework supports dense tensors, and we leave the sparse implementations for future releases. We use Python (v3.7.0) compiler and the OpenMPI (v2.1.2) library available on Grizzly.

\subsection{Data Generation}
We generate a synthetic tensor, say $\ten{A} \in \mathbb{R}^{n_{1},...n_{d}}$ , with known tensor-train ranks $r_{1},\dots,r_{d-1}$, and selected dimensions $n_{1},....n_{d}$. Each of the tensor train factors, e.g., $\ten[][i]{G}$, gets elements sampled from a uniform distribution between 0 and 1. The tensor $\ten{A}$ is then generated as a product of the TT factors and is distributed among the processors if its size is too large. Specifically, if the tensor is too large, we first generate the TT cores as described above, reshape them into matrices, and distribute them along the 1D processor grid. We then perform a distributed matrix multiplication of the factors and repeat the process until we obtain the final unfolded tensor. Then, we apply distributed reshape operation on the unfolded tensor to obtain the desired $d$-dimensional tensor. 

\subsection{Scalability}
We evaluate the scalability of the distributed tensor train algorithm looking at both strong and weak scaling. We also analyze the scaling performance with respect to tensor train ranks. We used 16, 32, 64, 128, and 256 processors/cores for these experiments. For the scaling experiments, we collect the total time taken for the nonnegative matrix factorization for all $d-1$ factors, having fixed the number of NMF iterations to 100. This is done ten times and the average times are reported. We also report the time spent on compute, communication, and I/O operations in the TT algorithm. The computation costs are: \emph{GR}---the  local  computations  of a Gram  matrix  ($\mat{W^{T}}\mat{W}$ or $\mat{H}\mat{H^{T}}$),  which  is  of  size  $r \times r$ ; \emph{MM}—matrix-matrix  multiplications  using  local  (MPI  rank  specific)  input  matrix  and  factor  matrices; \emph{MAD}---element-wise matrix multiplication and division operations; \emph{Norm}---l2 norm computation of local matrices; \emph{INIT}---initialization of factor matrices. The communication cost includes{:} all\textunderscore gather (\emph{AG})---time taken for global matrix-matrix multiplications while distributing the results across all processors; all\textunderscore reduce (\emph{AR})---the time required to compute global Gram matrices, and reduce-scatter (\emph{RSC})---the time to compute global matrix-matrix product using a reduce-scatter operation.
 We also report the breakdown of the scaling performance for distributed reshaping and I/O operations along with NMF operations.

\subsubsection{Strong Scaling}
\begin{figure*}
\centering
\begin{subfigure}{.33\textwidth}
  \centering
  \includegraphics[width=\linewidth]{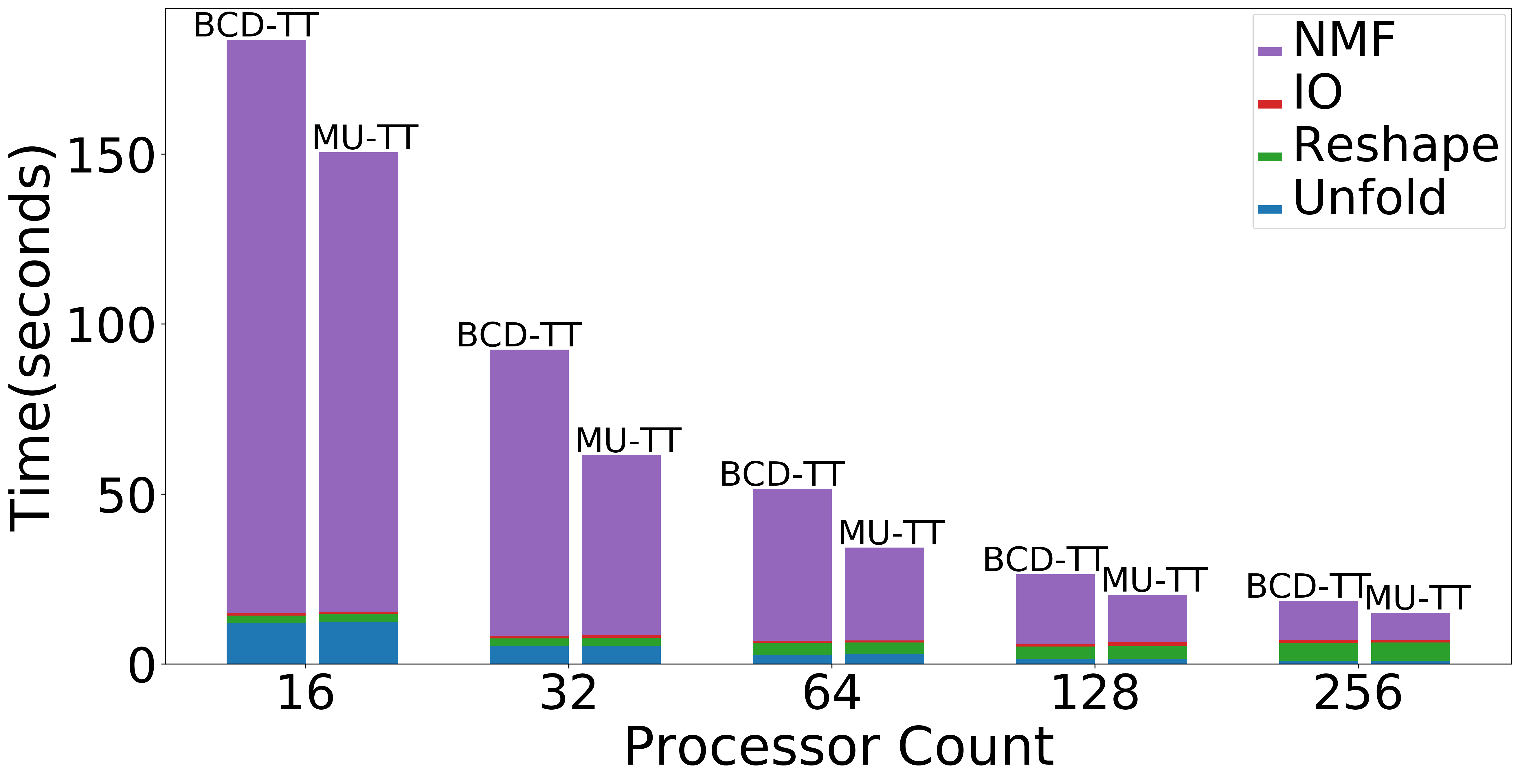}
  \caption{Strong scaling (Overall)}
\end{subfigure}%
\begin{subfigure}{.33\textwidth}
  \centering
  \includegraphics[width=\linewidth]{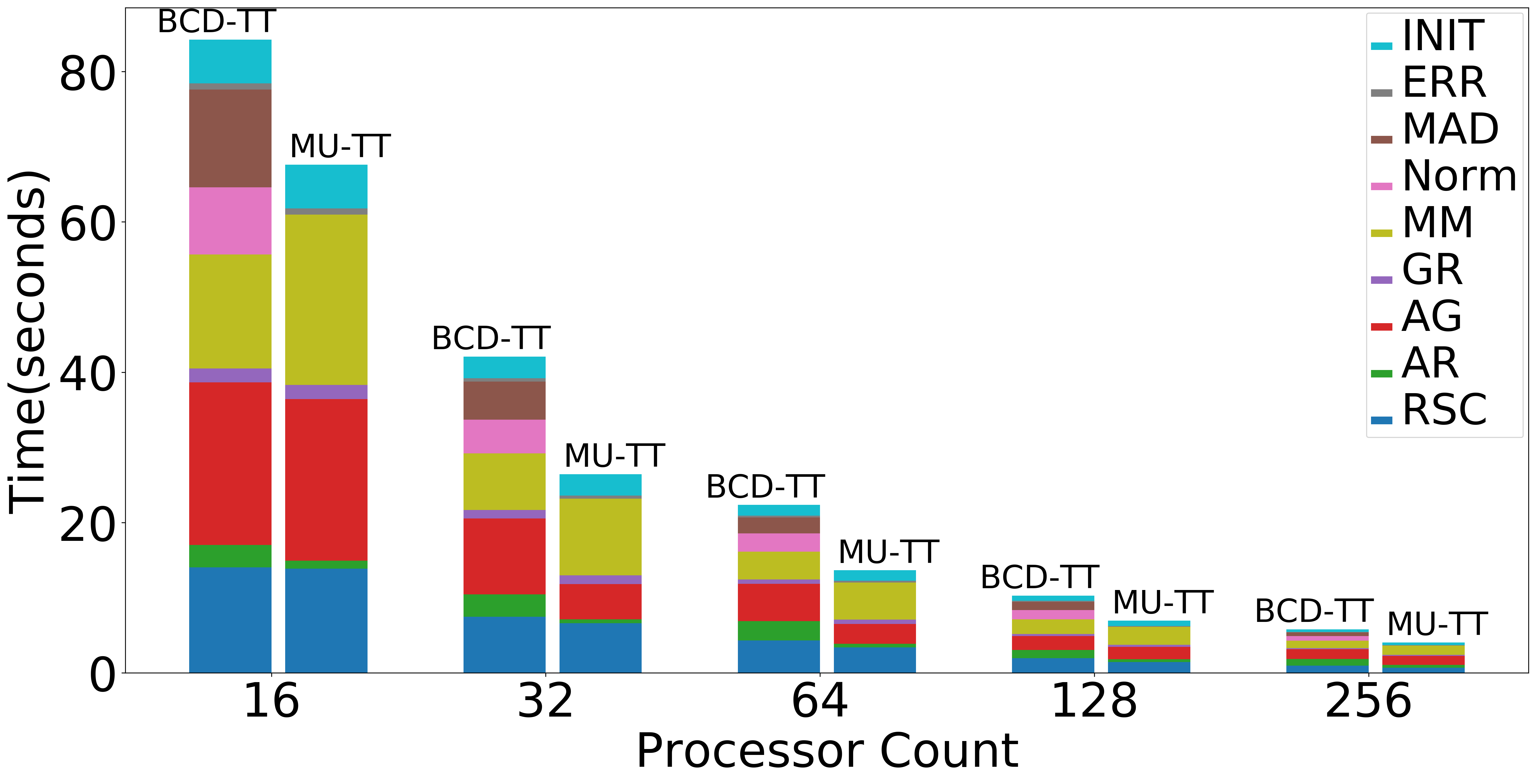}
  \caption{Strong scaling (NMF)}
\end{subfigure}
\begin{subfigure}{.33\textwidth}
  \centering
  \includegraphics[width=\linewidth]{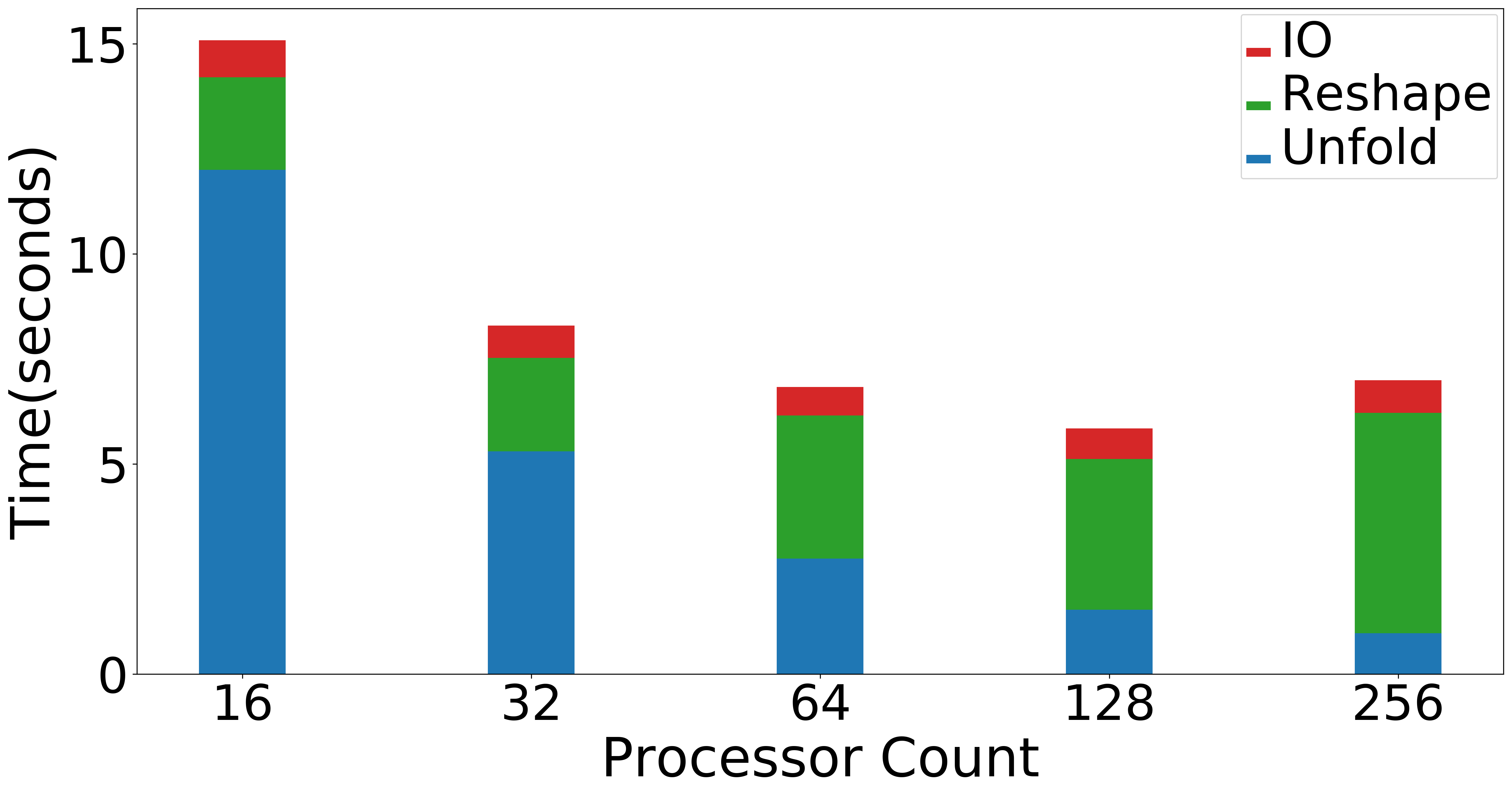}
  \caption{Strong scaling (Data Operations)}
\end{subfigure}

\caption{Strong scaling experiments }
\label{fig:strong}
\end{figure*}

We use data with fixed size to be $256\times256\times256\times256$ (i.e. 16GB) and vary the number of processors that we use to compute the distributed tensor train decomposition. The times taken for NMF of all factors, for data operations, and the overall TT time for two NMF algorithms, the BCD (block coordinate descent) and the MU (multiplicative update) times, across processor grid of sizes $2^{k} \times 2 \times 2  \times 2$, where $1 \leq k \leq 5$, are presented in Figure~\ref{fig:strong}.

The strong scaling experiments are run with  tensor-train ranks (TT-ranks) set to be 1,10,10,10, and 1 in the respective dimensions. The scaling results show that the overall TT performance achieves better FLOPS with larger grid size as the overall running time decreases with larger processor size. The scaling saturates at larger core sizes due to the inter-processor communication and the smaller matrix operations within the local computation kernels.

\subsubsection{Weak Scaling}
\begin{figure*}
\centering
\begin{subfigure}{.33\textwidth}
  \centering
  \includegraphics[width=\linewidth]{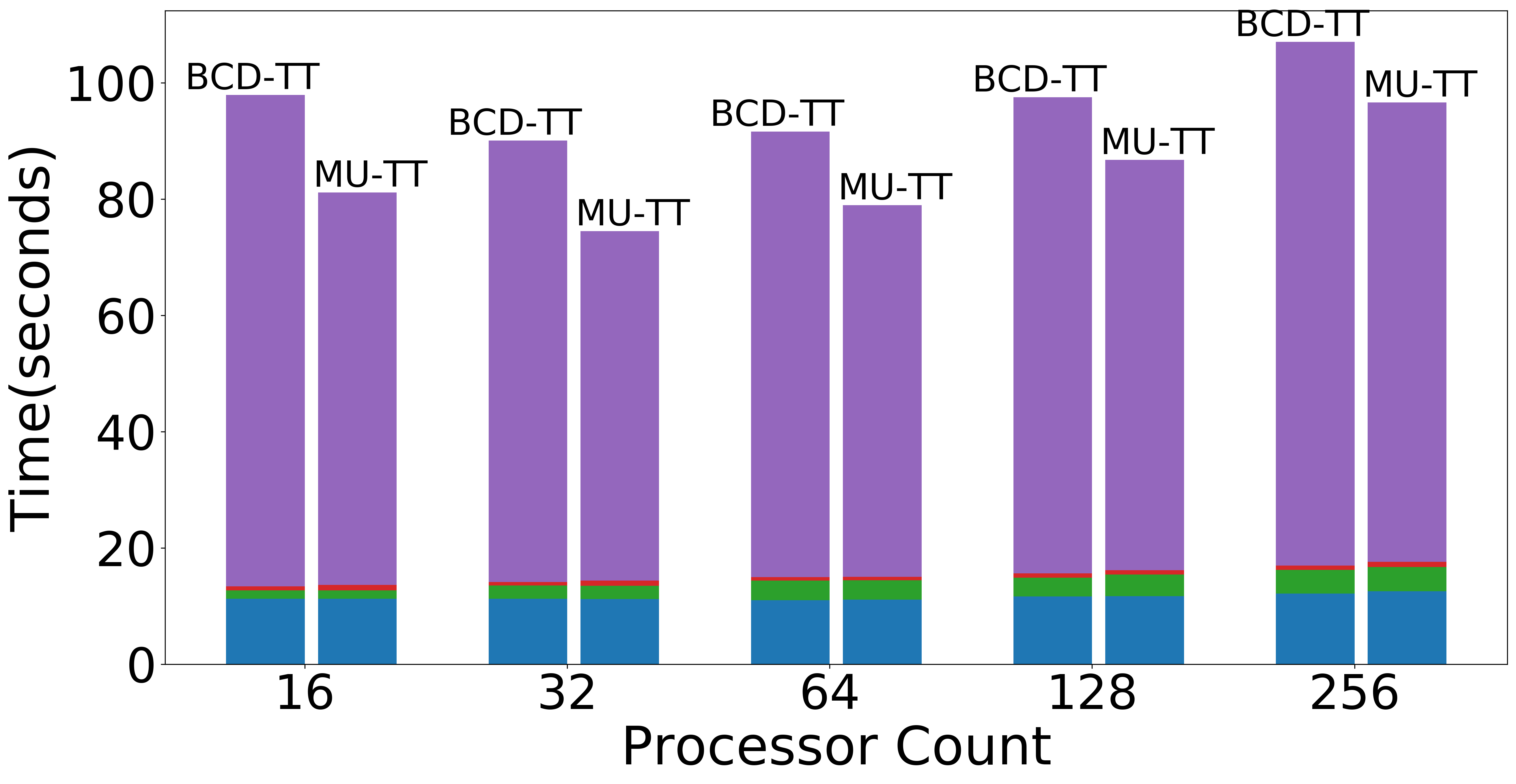}
  \caption{Weak scaling (Overall)}
\end{subfigure}%
\begin{subfigure}{.33\textwidth}
  \centering
  \includegraphics[width=\linewidth]{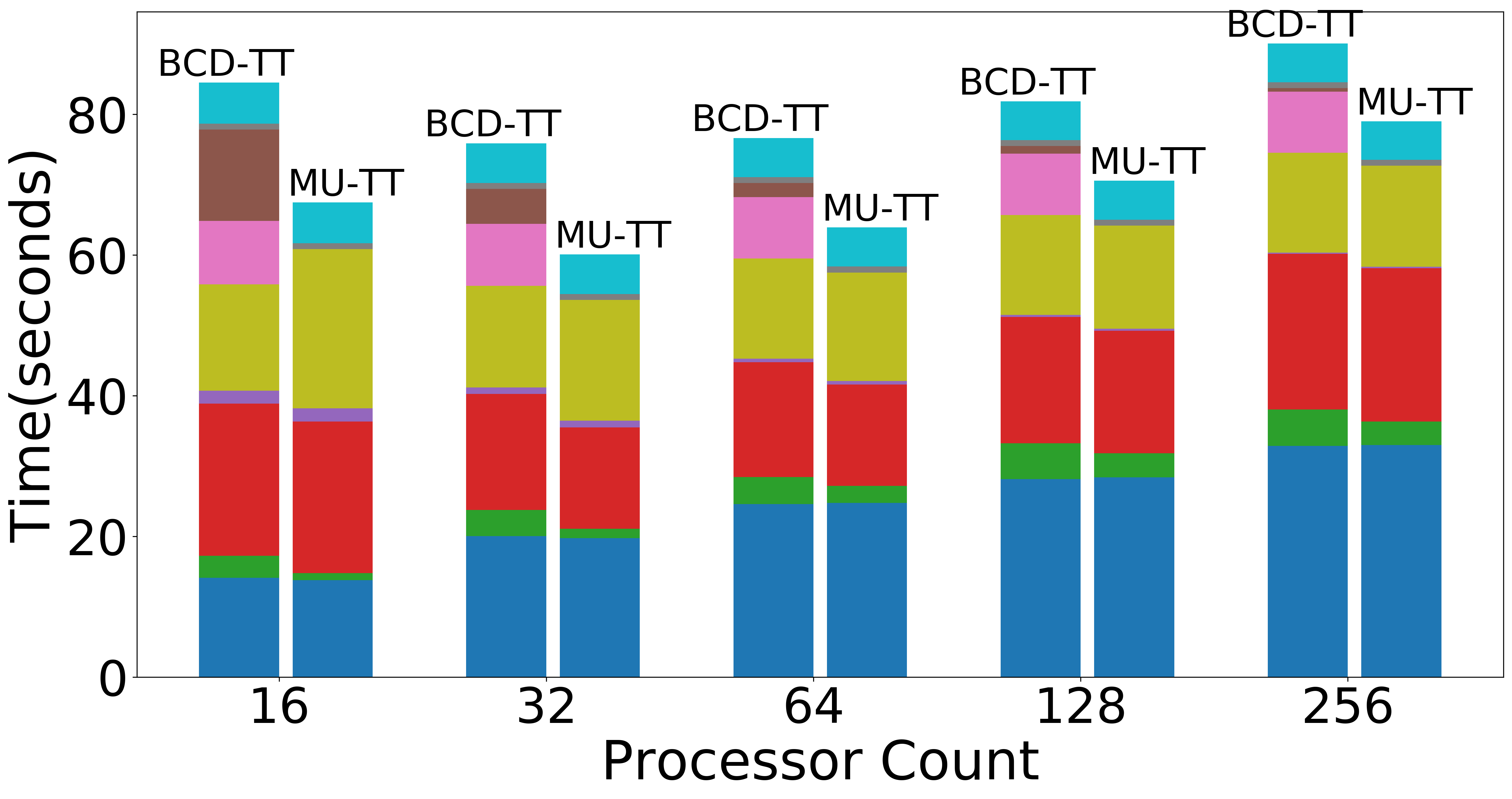}
  \caption{Weak scaling (NMF)}
\end{subfigure}
\begin{subfigure}{.33\textwidth}
  \centering
  \includegraphics[width=\linewidth]{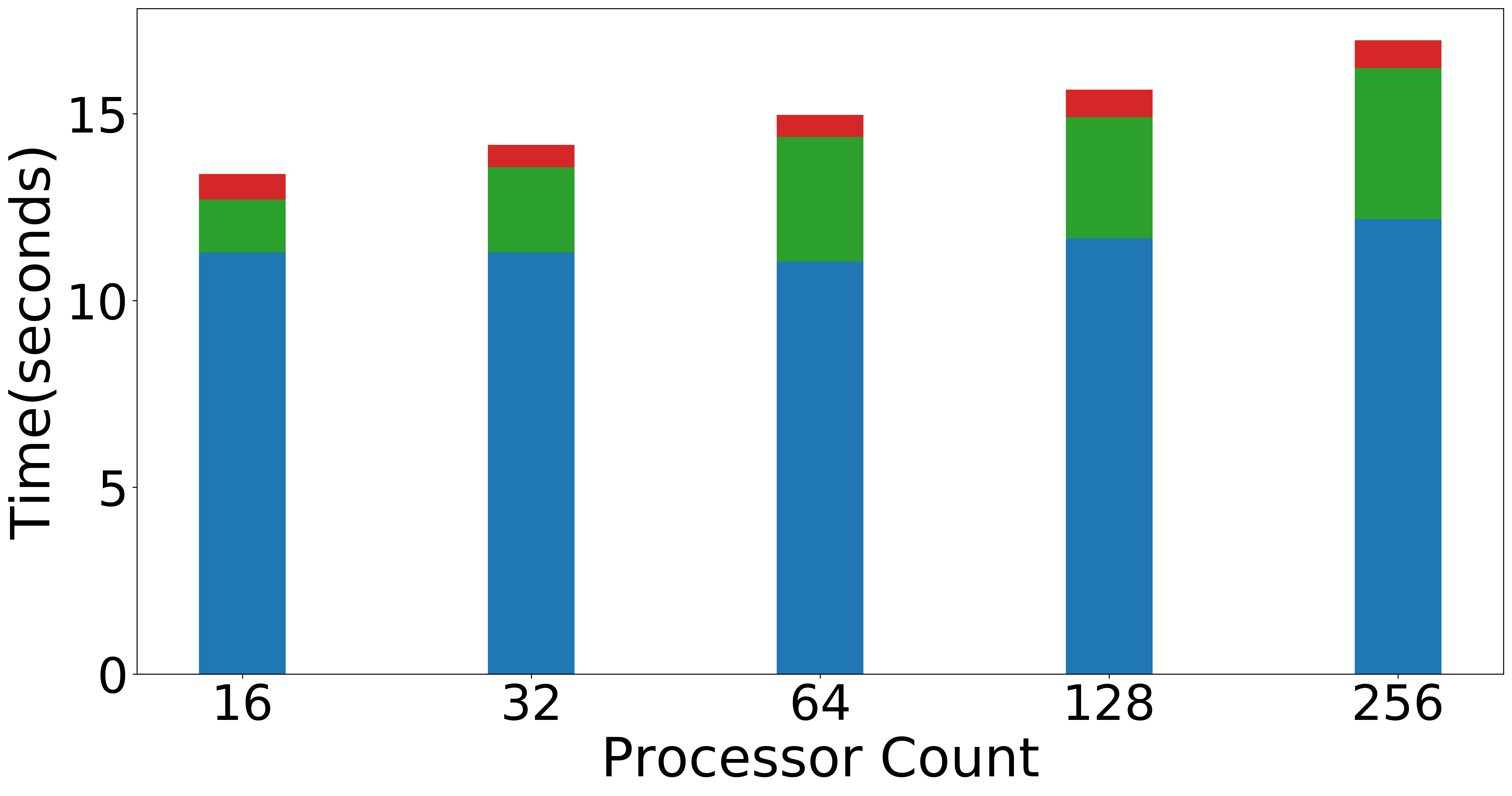}
  \caption{Weak scaling (Data Operations)}
\end{subfigure}

\caption{Weak scaling experiments}
\label{fig:weak}
\end{figure*}

To test the weak scaling performances of the proposed TT implementation, we report the performance per core for 100 iterations of each TT decomposition stage. Figure~\ref{fig:weak} illustrates the weak scaling performance for the proposed framework.

\begin{figure*}
\centering
\begin{subfigure}{.33\textwidth}
  \centering
  \includegraphics[width=\linewidth]{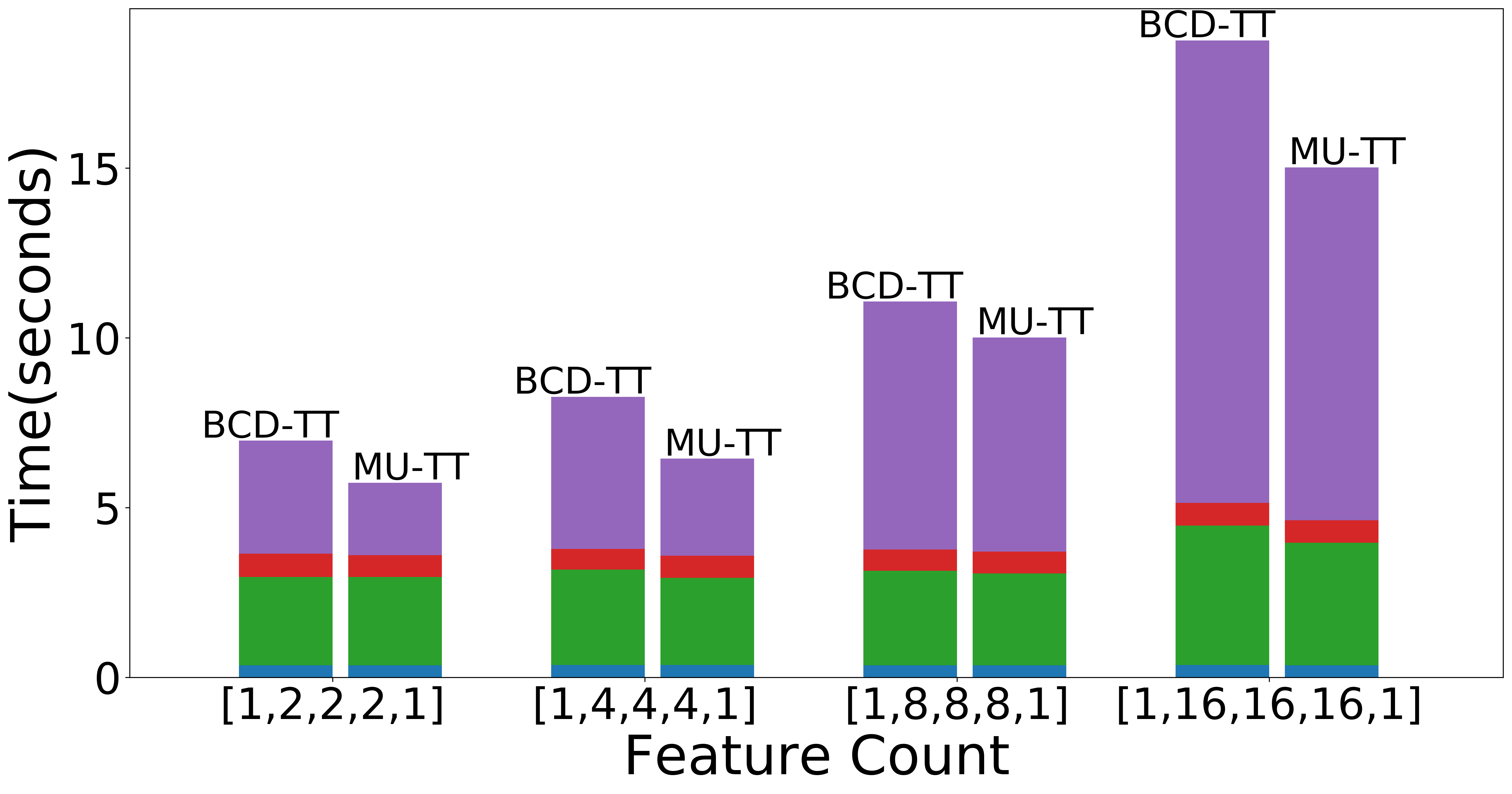}
  \caption{Scaling with TT-ranks (Overall)}
\end{subfigure}%
\begin{subfigure}{.33\textwidth}
  \centering
  \includegraphics[width=\linewidth]{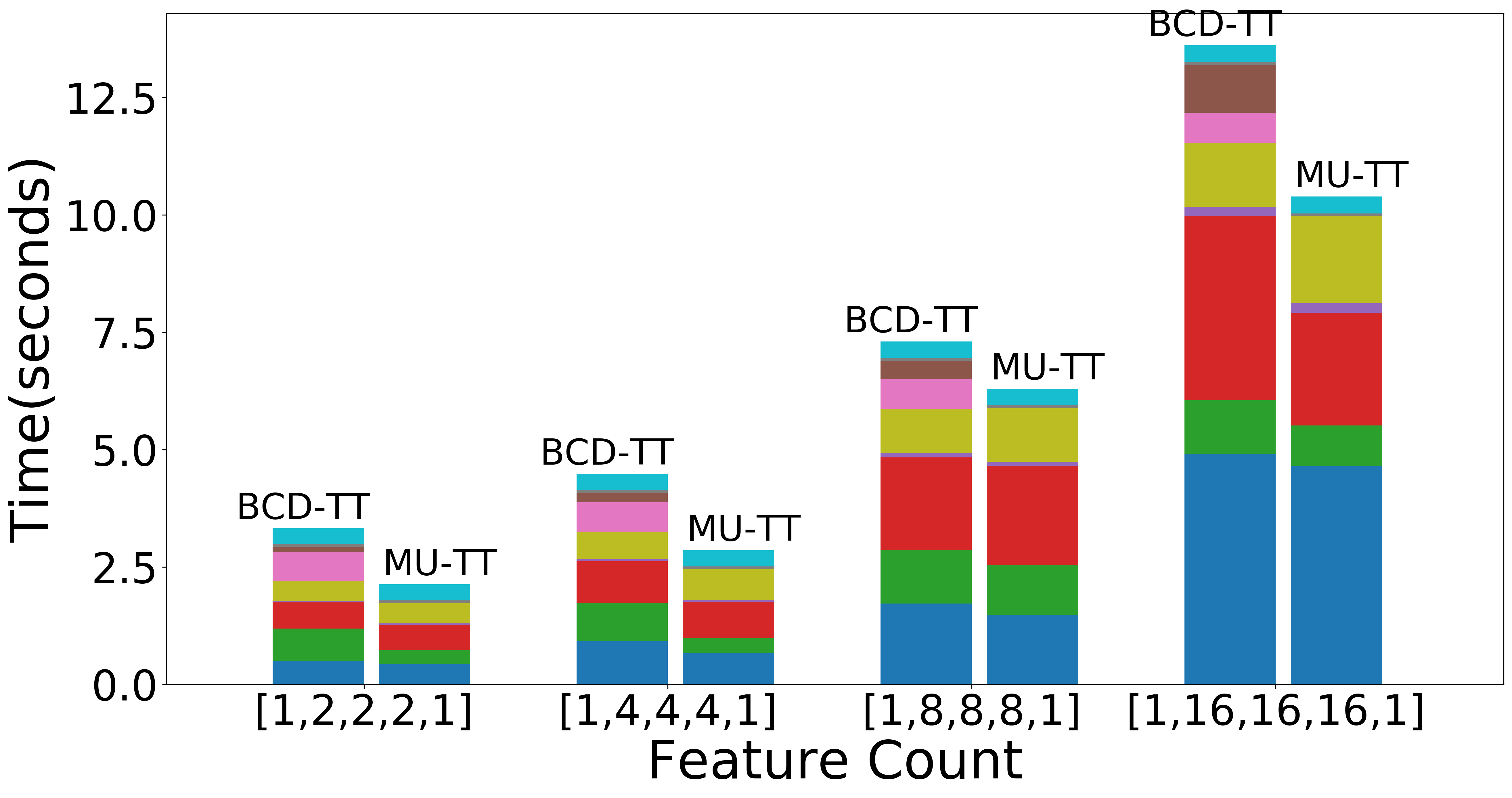}
  \caption{Scaling with TT-ranks (NMF)}
\end{subfigure}
\begin{subfigure}{.33\textwidth}
  \centering
  \includegraphics[width=\linewidth]{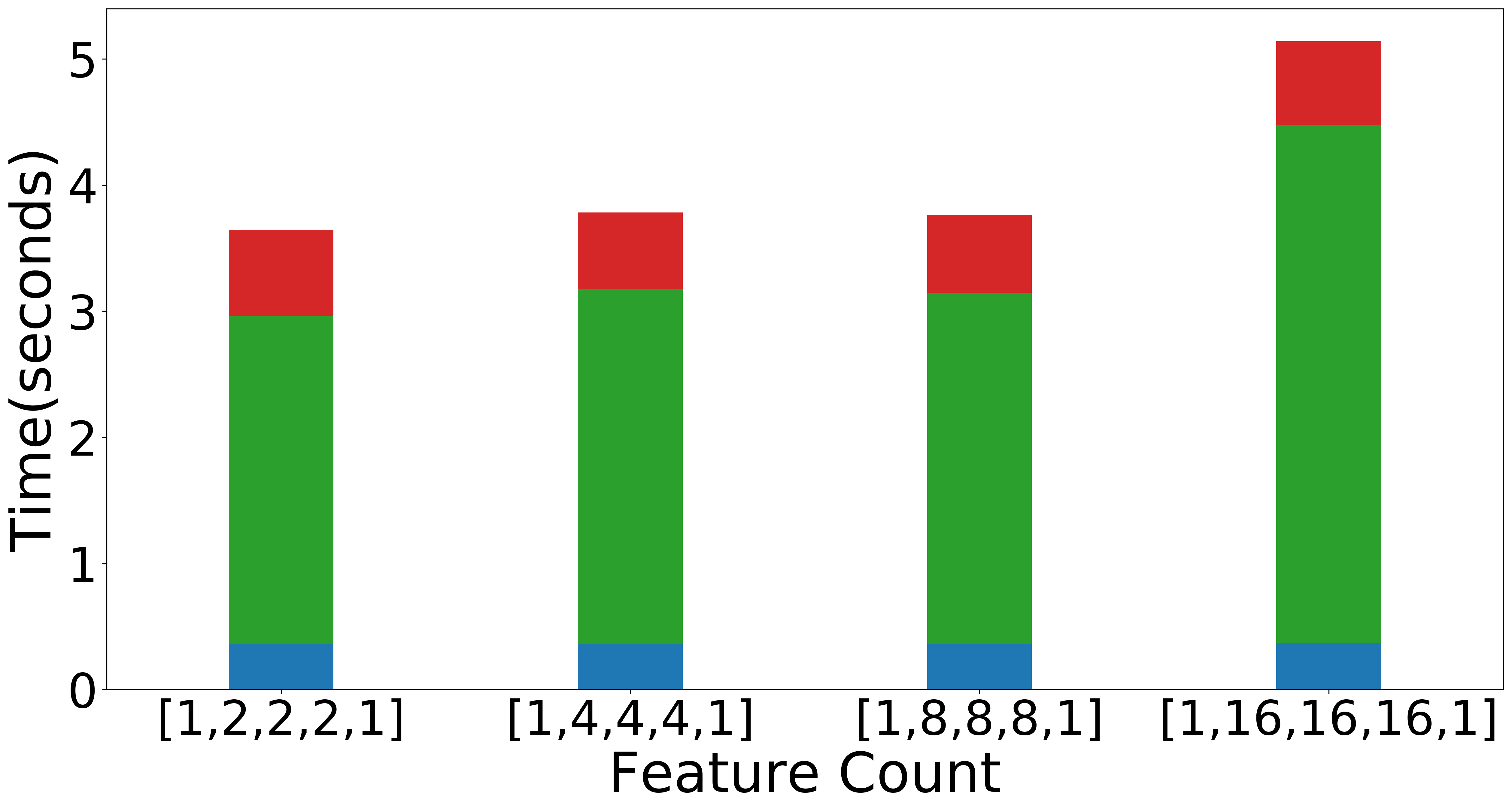}
  \caption{Scaling with TT-ranks (Data Operations)}
\end{subfigure}

\caption{Scaling experiments with respect to TT-ranks }
\label{fig:features}
\end{figure*}

For this experiment, the size of data is fixed per processor, while we scale up the processor  and the data sizes by the same factors.  Similar to the strong scaling, we use processor grid of sizes $2^{k} \times 2 \times 2  \times 2$, where $1 \leq k \leq 5$. In addition to that, we vary the data size as $256^{k} \times 256 \times 256  \times 256$, where $1 \leq k \leq 5$. The condition for $k=1$ matches with the $k=1$ setup in strong scaling. The size of data varies from 16GB to 256GB for a processor count ranging from 16 to 256. Again, the scaling performance degrades slightly for larger processors and data sizes due to the inter-nodal communications and the I/O involved. 
\begin{figure*}%
    \centering
    \begin{subfigure}{.31\textwidth}
    \includegraphics[width=\linewidth]{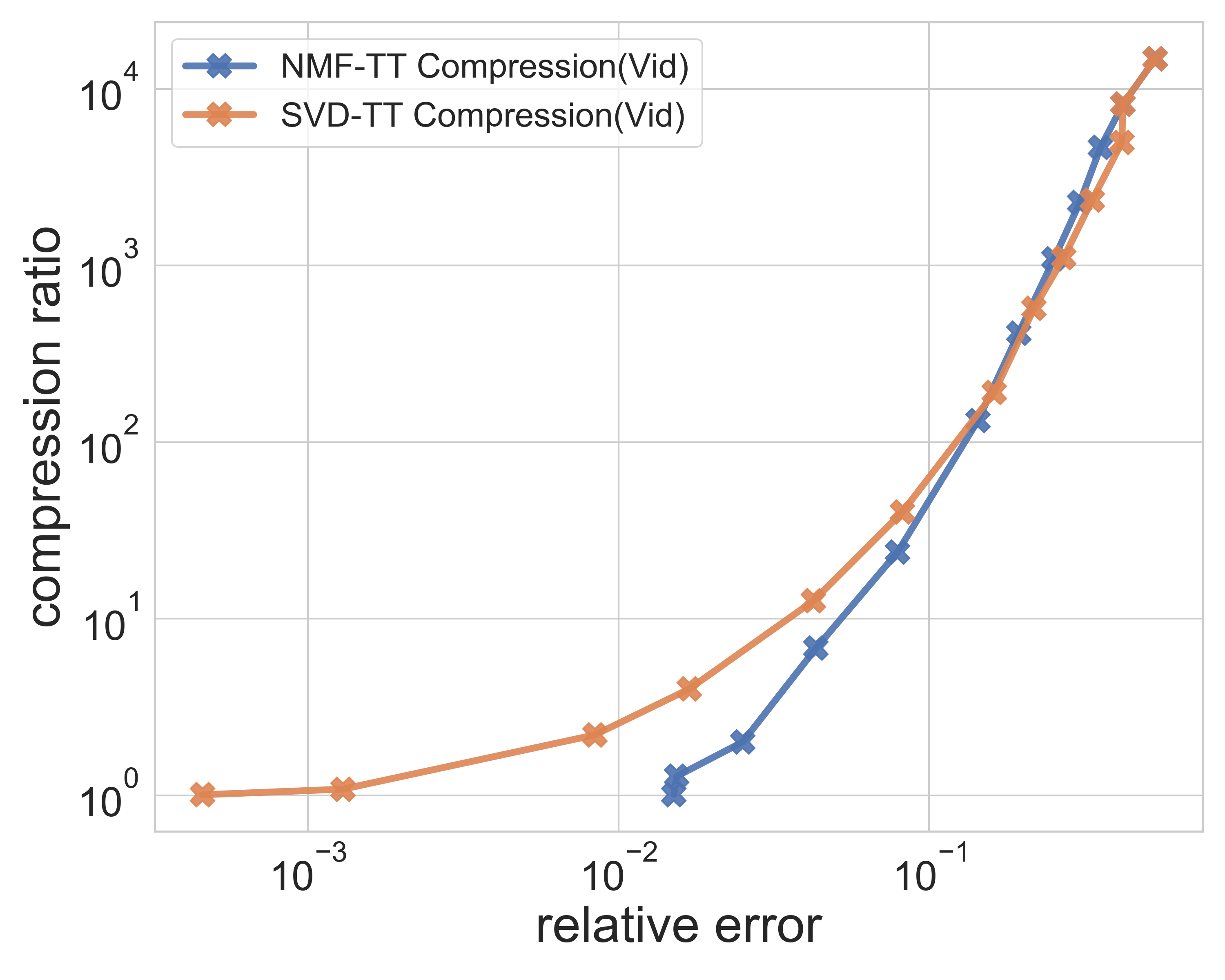}
    \caption{ } \label{fig:yale}
    \end{subfigure}
    \qquad
    \begin{subfigure}{.31\textwidth}
    \includegraphics[width=\linewidth]{images/compression_vs_error_vid_svd_nmf_final_loglog.png}
    \caption{ } \label{fig:vid}
    \end{subfigure}
    \begin{subfigure}{.31\textwidth}
    \qquad
    \includegraphics[width=\linewidth]{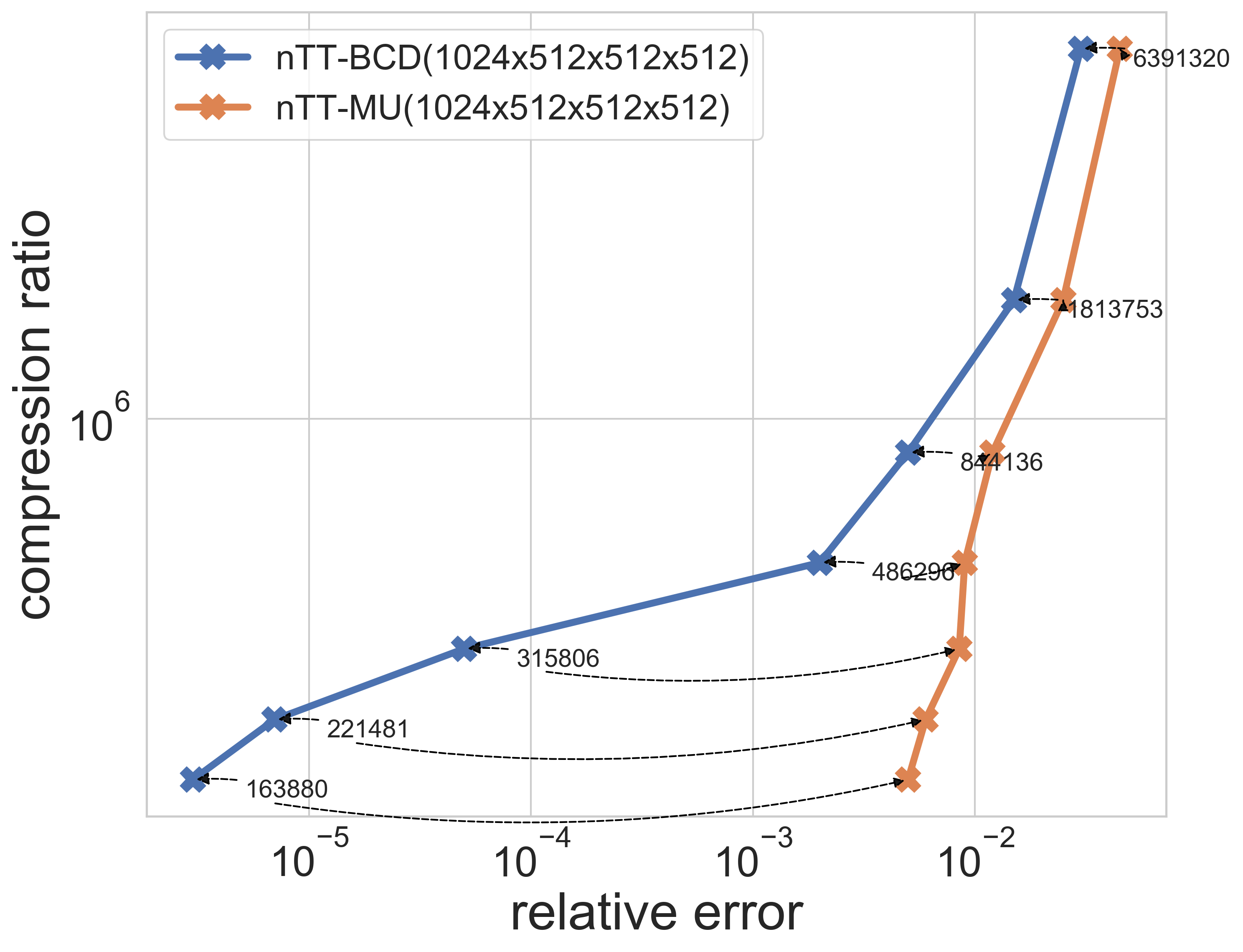}
    \caption{ } \label{fig:synthetic}
    \end{subfigure}
    \caption{Compression ratio vs relative error with tensor train decomposition on datasets a)Yale face b) Video c) Synthetic data (500GB)}%
    \label{fig:TT}%
\end{figure*}
\subsubsection{Scaling with TT ranks}
To demonstrate the scalability of our framework for different tensor train ranks, we fix the total number of processors to be 256 for a tensor of size $256 \times 256 \times 256  \times 256$, and vary the inner tensor train ranks to analyze the effect of $r$ on scaling. Figure \ref{fig:features} shows the scaling with TT ranks results, where each TT rank $r$ is varied in $\{2,4,8,16\}$.

\subsection{Application to real-world dataset}

\subsubsection{Data Description}
\subsubsubsection{Extended Yale Face Dataset B} We first demonstrate the compressibility on Extended Yale Face Dataset B \cite{GeBeKr01,wang2019principal,wang2017efficient} that includes 38 people with 9 poses under 64 illumination conditions. Each image from the Yale Face dataset has size of $192 \times 168$, where each image is down-sampled to $48 \times 42$ for comparison with an existing method \cite{GeBeKr01}. The formatted 4D tensor dimensions are $48 \times 42 \times 64 \times 38$.  We also demonstrate denoising on the same dataset by adding Gaussian noise $N(0,900)$ to each voxel of the tensor.

\begin{figure}
\centering
\begin{subfigure}[b] {0.45\textwidth}
   \includegraphics[width=0.9\linewidth]{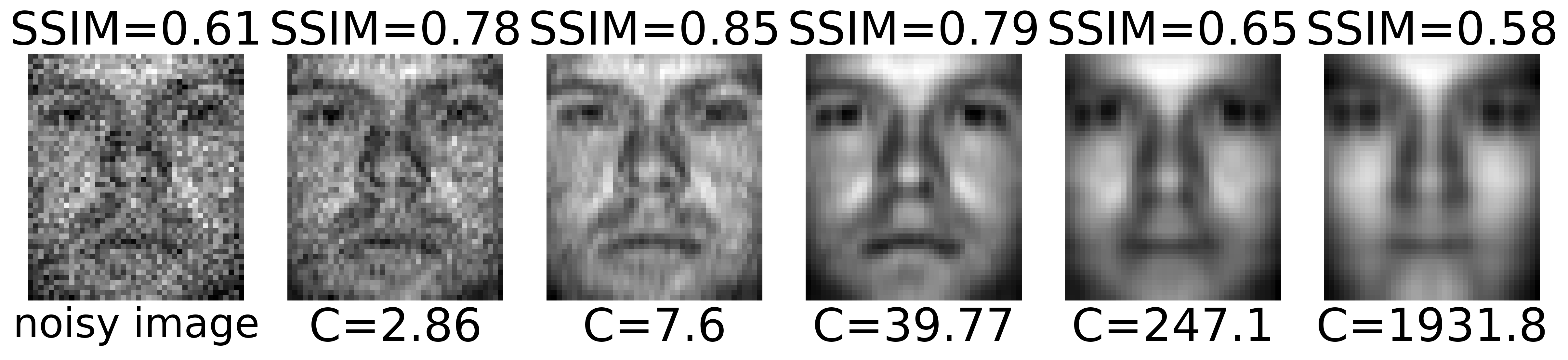}
   \caption{}
   \label{fig:Ng1} 
\end{subfigure}

\begin{subfigure}[b]{0.45\textwidth}
   \includegraphics[width=0.9\linewidth]{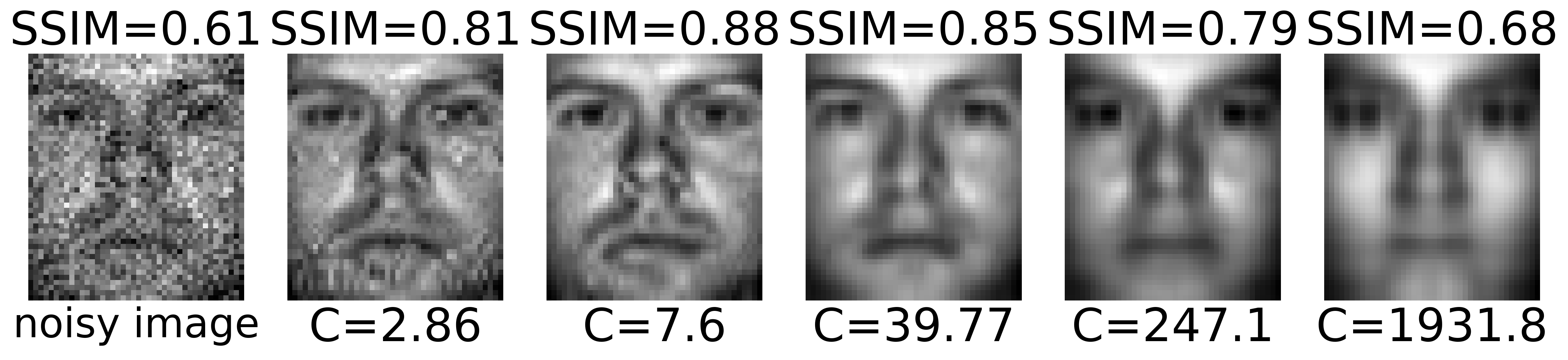}
   \caption{}
   \label{fig:Ng2}
\end{subfigure}
\caption{Demonstration of image denoising. (a) Denoising with SVD-TT (b) Denoising with NMF-TT } \label{fig:denoise}
\end{figure}

\subsubsubsection{Video} 
 The video tensor ($100 \times 260 \times 3 \times 85$) obtained from a high-speed camera video for gun shooting \cite{youtube}, comprises 4 dimensions, where the first two dimensions correspond to a monochromatic image, the third dimension is the channel, and the fourth one is the frame count.

\subsubsection{Compression Ratio and Reconstruction Error}
For a $d$-way data tensor $\ten{A}$ of size $n_{1},n_{2},...,n_{d}$, the reconstructed tensor from the tensor train factors $\ten[i]{G}$ is given by $\ten{\tilde A}$. Then the relative error $\epsilon$ for reconstruction is given as 
\begin{equation}
    \epsilon= \frac{\Vert \ten{A} - \ten{\tilde A}\Vert_F}{\Vert \ten{A} \Vert_F}.
\end{equation}
 
If the tensor train ranks of the decomposition are $r_{0}= r_d = 1,$ $r_{i} \geq 1$ for $1 \leq i < d$, then the compression ratio is measured as 
\begin{equation}
    C = \prod_{i=1}^{d} n_i /(\sum_{i=1}^{d} n_{i}*r_{i-1}*r_{i}) 
\end{equation}

Each data point in the Figures~ \ref{fig:yale} and \ref{fig:vid} is the compression ratio $C$ at the desired relative error $\epsilon$ at each TT decomposition stage. The targeted relative errors at each TT stage for selection of $r_i$ was set to 0.5, 0.25, 0.125, 0.075, 0.01, 0.005 and 0.001. Lower TT rank produces higher compression and higher reconstruction error, whereas higher TT rank produces lower compression and better reconstruction. For the Yale face dataset, the compression ratio $C$ varies from 1.13 with $\epsilon= 0.04$ to $C$ of 2.55e4 for $\epsilon= 0.55$ with  nTT. Similarly with TT, the compression ratio $C$ varies from 1.32 with $\epsilon= 0.013$ to $C$ of 2.55e4 for $\epsilon= 0.55$. Also, for the video dataset, the compression ratio $C$ varies from 1.01 with $\epsilon= 0.015$ to $C$ of 1.47e4 for $\epsilon= 0.54$ with  nTT. Similarly with TT, the compression ratio $C$ varies from 1.007 with $\epsilon= 0.0004$ to $C$ of 1.47e4 for $\epsilon=  0.53$.

\subsubsection{Application to image denoising}

To demonstrate the efficacy of non-negative tensor train over the regular tensor train, we apply both techniques for the decomposition of a noisy Yale Face sample and report the denoising performance. The metric that we choose to evaluate the correctness of reconstruction compared to the noise-free sample is the structural similarity (SSIM) index~\cite{clifton1938osteopetrosis}. SSIM is a widely used metric for image similarity measures in computer vision applications. SSIM ranges on a scale of [0,1], where 1 is the best match.  In Figure~\ref{fig:denoise}, we report the SSIM values of the reconstructed images with respect to the noiseless ground truth image. Figure~\ref{fig:Ng1} corresponds to the regular tensor train based reconstruction, whereas Figure~\ref{fig:Ng2} corresponds to the non-negative TT based reconstruction. For both of these figures, the images from left to right correspond to the reconstructed images with decreasing TT-ranks and increasing compression rates. The value of the top of each image corresponds to the SSIM measure with respect to the ground truth and the value on the bottom correspond to the compression rate. We can observe that increasing the compression-rate-based decomposition eliminates noise significantly and results in the reconstructed image to resemble the original noise-free image. For regular TT/SVD-TT based reconstruction, the best SSIM reported for the reconstructed image is 0.85 whereas with the non-negative TT/nTT based reconstruction, the best SSIM reported is 0.88. For given TT ranks, the reconstructed image SSIM for nTT is better than that for the TT.

\subsubsection{Compression of large synthetic data}

We demonstrate the compression ratios with lower reconstruction error for a 500 Gigabyte (GB) matrix with dimensions $1024 \times 512 \times 512 \times 512$ and tensor train ranks =[1,20,30,40,1]. We synthetically generate the data in a distributed manner as discussed in the data generation section. Figure~\ref{fig:synthetic} shows compression ratios with two different NMF optimization methods, BCD vs multiplicative update, both of which are based on minimization of the Frobenius norm. For the BCD optimization-based NTF, the compression ratio $C$ varies from 163880 with $\epsilon=3e-6$ to $C$ of 6391320 for $\epsilon= 0.03$ with non-negative tensor train. Similarly, with multiplicative update algorithm based NTF, the relative error $\epsilon$  varies from 0.005 to 0.045 for the same compression range. This experiment demonstrates the tradeoff between two different NMF update algorithms BCD and MU for nTT. As per the scaling plots \ref{fig:strong},\ref{fig:weak} and \ref{fig:features}, MU algorithm demonstrates better timings. However, BCD achieves a better compression rate with lower reconstruction error compared to MU as per Figure \ref{fig:synthetic}. 

\section{Conclusion and Future directions}
Here, we introduce a distributed non-negative tensor train, nTT, algorithm that is capable of computing tensor train to a prescribed relative error. We demonstrate the nTT scaling performance on synthetic data, and establish its ability to decompose a 500GB tensor. Finally, we apply the algorithm to various real and synthetic datasets to demonstrate the nTT data compression capabilities. In the future, we aim to apply our framework to large real-world datasets such as seismic datasets, satellite images, medical images, etc., in order to have an efficient highly compressed representation and to be able to do inference/classification from the nTT low-dimensional representations. 

\section{Acknowledgements}
This research was funded by Laboratory Directed Research and Development  (20190020DR), and resources were provided by the Los Alamos National Laboratory Institutional Computing Program, supported by the U.S. Department of Energy National Nuclear Security Administration under Contract No. 89233218CNA000001.

\bibliographystyle{IEEEtran} 
\bibliography{references}
\end{document}